\documentclass[]{aa}  

\usepackage{graphicx}
\usepackage{savesym}
\usepackage{amsmath}
\savesymbol{iint}
\savesymbol{iiint}
\usepackage{txfonts}
\restoresymbol{TXF}{iint}
\restoresymbol{TXF}{iiint}

\usepackage[T1]{fontenc}
\usepackage{ae,aecompl}

\usepackage{subfig}
\usepackage{multirow}
\usepackage{booktabs}
\usepackage{rotating}
\usepackage{natbib}
\begin{document}

   \title{The asteroseismic potential of {\it CHEOPS}}


   \author{A. Moya
          \inst{1,2,3}\fnmsep
          \and
          S. Barcel\'o Forteza
          \inst{4}
          \and
          A. Bonfanti
          \inst{5}
          \and
          S.J.A.J. Salmon
          \inst{5}
          \and
          V. Van Grootel
          \inst{5}
          \and
          D. Barrado
          \inst{4}
          }

   \institute{School of Physics and Astronomy, University of Birmingham, Edgbaston, Birmingham, B15 2TT, UK
              \email{A.Moya@bham.ac.uk}
         \and
             Stellar Astrophysics Centre, Department of Physics and Astronomy, Aarhus University, Ny Munkegade 120, DK-8000 Aarhus C, Denmark
		 \and
         Dept. F\'isica Te\'orica y del Cosmos. Universidad de Granada. Campus de Fuentenueva s/n. 10871. Granada. Spain.
         \and
         Depto. Astrof\'isica, Centro de Astrobiolog\'ia (INTA-CSIC), ESAC campus, Camino Bajo del Castillo s/n, 28692, Villanueva de la Ca\~nada, Spain
         \and
         Space sciences, Technologies and Astrophysics Research (STAR) Institute, Universit\'e de Li\`ege, All\'ee du Six Ao\^ut 19C,\\ B-4000, Li\`ege, Belgium
             }

   \date{Received September 15, 1996; accepted March 16, 1997}

 \abstract
 {Asteroseismology has been impressively boosted during the last decade mainly thanks to space missions such as {\it Kepler/K2} and {\it CoRoT}. This has a large impact, in particular, in exoplanetary sciences since the accurate characterization of the exoplanets is convoluted in most cases with the characterization of their hosting star. Until the expected launch of the ESA mission {\it PLATO 2.0}, there is almost a decade where only two important missions will provide short-cadence high-precision photometric time-series: NASA--{\it TESS} and ESA--{\it CHEOPS} missions, both having high capabilities for exoplanetary sciences.}
 {In this work we want to explore the asteroseismic potential of {\it CHEOPS} time-series.}
 {Following the works done for estimating the asteroseismic potential of {\it Kepler} and {\it TESS}, we have analyzed the probability of detecting solar-like pulsations using {\it CHEOPS} light-curves. Since {\it CHEOPS} will collect runs with observational times from hours up to a few days, we have analyzed the accuracy and precision we can obtain for the estimation of $\nu_{\rm max}$, the only asteroseismic observable we can recover using {\it CHEOPS} observations. Finally, we have analyzed the impact of knowing $\nu_{\rm max}$ in the characterization of exoplanet host stars.}
 {Using {\it CHEOPS} light-curves with the expected observational times we can determine $\nu_{\rm max}$ for massive G and F-type stars from late Main Sequence on, and for F, G, and K-type stars from post-Main Sequence on with an uncertainty lower than a 5$\%$. For magnitudes V<12 and observational times from eight hours up to two days, the HR zone of potential detectability changes. The determination of $\nu_{\rm max}$ leads to an internal age uncertainty reduction in the characterization of exoplanet host stars from 52$\%$ to 38$\%$; mass uncertainty reduction from 2.1$\%$ to 1.8$\%$; radius uncertainty reduction from 1.8$\%$ to 1.6$\%$; density uncertainty reduction from 5.6\% to 4.7\%, in our best scenarios.}
 {} 
 


   \maketitle
%

\section{Introduction}

{\it CHEOPS} \citep{Fortier14} is the first small European Space Agency mission (ESA S-mission). Its launch is expected at the end of 2018 and its main scientific goal is the accurate characterization of transiting exoplanetary systems.

{\it CHEOPS} will collect high precision photometry, of the order of parts-per-million (ppm), depending on the stellar magnitude, in short cadence (1 minute). These precision and cadence are similar to those reached by {\it Kepler} \citep{Kepler} and {\it CoRoT} \citep{Corot}, but in the case of {\it CHEOPS} for brighter stars than with {\it Kepler} since its telescope aperture is significantly smaller (see Table \ref{tab:comp}).

\begin{sidewaystable*}
	\vspace{70mm}
	\caption{Comparison among different past, current, and future space missions with asteroseismic capabilities.}
	\centering
	\begin{tabular}{cccccccc}
		\hline
		& {\it MOST} & {\it CoRoT} & \multicolumn{2}{c}{\it Kepler}  & {\it BRITE} & {\it TESS} & {\it CHEOPS}\\ \hline
		& & & ${\it Kepler}$ & $K2$ & & &\\
		\cmidrule(lr){4-5}
		Mission years   & $\geq$13 & 6.5 & 4 & $\geq$4 & $\geq$4 & $\geq$2 & $\geq$3.5\\
		Telescope aperture (cm) & 15 & 27 & \multicolumn{2}{c} {95} & 5 $\times$ 3 & 4 $\times$ 10.5 & 32\\
		\multirow{2}{1cm}{\centering Orbit} & Geocentric & Geocentric & \multicolumn{2}{c}{Earth-trailing} & Geocentric  & High Earth & Geocentric\\
		& polar & polar & \multicolumn{2}{c}{heliocentric} & Low Earth & elliptical & sun-synchronous\\
		Duty cycle (in $\%$) & Variable & >90 & \multicolumn{2}{c}{>95} & Variable  & > 95 & [60, 100]  \\
		Cadence & <1 min & 32/512 s & \multicolumn{2}{c}{59 s/29 min} & 1 s  & 20s/2/30 min & 60 s \\
		Pointing freedom & Free & Several fixed FOVs & Fixed FOV & Free & Ecliptic plane  & All sky monitoring & Free \\
		Sky coverage (in square degrees) & PIT\tablefootmark{b} & <600 & 100 & $\approx$ 80$\%$ ecliptic plane & PIT\tablefootmark{b} & Almost full\tablefootmark{a} & PIT\tablefootmark{b} \\
		\hline
	\end{tabular}
	\tablefoottext{a}{Weak coverage for Dec in the range [-6$^\circ$,6$^\circ$].}
	\tablefoottext{b}{PIT = Pointing Individual Targets}
	\label{tab:comp}
\end{sidewaystable*}

With this work, we aim to explore the potential of {\it CHEOPS} for asteroseismic studies of exoplanet host stars. There are a number of studies in the literature showing the benefit of obtaining asteroseismic observables of exoplanet host stars \citep{Bazot05,Vauclair08,Escobar12,Moya13,Huber13,Lillo14,Silva15,Davies16,Huber18}. These additional observables allow a more precise determination of stellar mass, radius, and age, having a direct impact on the understanding of the exoplanetary system. Although in recent years the number of techniques for characterizing exoplanets independently of the stellar generic physical values has increased, in general the exoplanet observations are convoluted with the star, and characterizing the planets means to deconvolute the stellar contribution from the observables. On the other hand, the age of the system is critical from an astrobiological point of view and to model and explain the dynamical evolution of exoplanetary systems. Evaluating whether an exoplanet has had enough time for developing detectable biomarkers can only be done by knowing the age of the planet. Nowadays, the only way of dating an exoplanet is dating its hosting star \citep[see][for a review]{Barrado16}. An additional benefit of the asteroseismic studies of photometric time-series is the proper removal of these oscillations from the light-curve, allowing a more accurate modeling of the observing transits, mainly their ingress and egress.

Among the more than 20 known pulsational types along the HR-diagram \citep{Aerts10}, solar-like pulsations are one of the most well-known. These oscillations are produced by the convective zone near the surface, exciting stochastically high-order pressure modes in a broad frequency range. In principle, these solar-like oscillations are expected for all stars that posses a convective envelope, as FGK-type stars. Due to the special nature of solar-like oscillations (they pulsate in the frequency asymptotic regime), they provide in first approximation important information about general physical characteristics of the star, such as its mean density and/or its surface gravity. Another important pulsational type for characterizing exoplanet host stars are the so-called $\delta$~Scuti pulsations. $\delta$~Scuti stars are A-F classical pulsators excited by $\kappa$-mechanism \citep{Chevalier71} with frequencies around the fundamental radial mode, that is, at the middle of the stellar frequency spectrum range.

Following the works of \citet{Chaplin11} and \citet{Campante16} done in the context of {\it Kepler} and {\it TESS} \citep{Tess} space missions respectively, we have estimated the potential of {\it CHEOPS} for detecting solar-like pulsations. We have also studied the impact of different observational times and duty cycles in the accuracy\footnote{In this work we use the term "accuracy" when we can compare our predictions with reliable observations, such as $\nu_{\rm max}$ obtained using Kepler light curves, and "precision" when we show the internal dispersion of our estimations.} reached for the asteroseismic observables.

One of the main characteristics of {\it CHEOPS} is its short observational time per run, ranging from hours up to a few days in some cases. This makes it impossible to obtain individual frequencies. On the other hand, the frequency with the largest spectrum power (the so-called $\nu_{\rm max}$) is easier to obtain since its determination only needs the estimation of the Gaussian-like frequency power excess envelope. Therefore, we have focused on the potential observation of $\nu_{\rm max}$ using {\it CHEOPS} time series, which is a proxy for the stellar $\log{g}$ \citep{Brown91,Kjeldsen95}. We have also studied the potential of {\it CHEOPS} for observing $\nu_{\rm max}$ in the case of $\delta$~Scuti stars. This is the first time, up to our knowledge, that such a short time series from space are analyzed in an asteroseismic context and the reached precision studied.

The proper {\it CHEOPS} characteristics makes it a perfect space mission for the precise characterization of exoplanetary systems not covered by other past and current missions. In table \ref{tab:comp} we present a summary of the main characteristics of these past and current space missions with asteroseismic capabilities. The coverage of the ecliptic plane, its telescope aperture, and its pointing facilities make {\it CHEOPS} an unique opportunity for studying certain systems, thus complementing {\it TESS}, the other space mission with asteroseismic capabilities during the next decade, besides nano-satellite missions such as {\it BRITE} \citep{BRITE}. In particular, the zone of the sky not covered by {\it TESS} is the ecliptic plane, roughly in the range Dec = [-6$^\circ$, 6$^\circ$], where {\it CHEOPS} can do its best. In addition, {\it CHEOPS} can properly observe fainter stars than {\it TESS} due to its larger telescope aperture. For example, besides {\it CHEOPS} and {\it TESS} magnitudes are not fully equivalent, we can understand the difference comparing the shot noises reached by both missions for a $V$=12 and $I_c$=12 star respectively. In the case of {\it CHEOPS}, a $V$=12 star can be observed with a shot noise of 623 ppm (see table \ref{tab:sigma}), and {\it TESS} can observe a $I_c$=12 star with a shot noise around 1100 ppm \citep{Campante16}. Therefore, {\it TESS} and {\it CHEOPS} are complementary missions from an asteroseismic point of view.

In section \ref{sec:est_gen} we describe the procedure for estimating the solar-like pulsation-detection potential of {\it CHEOPS} and show the results obtained, the region in the HR-diagram where these stars are located, and the dependences of the boundaries of this region. In section \ref{sec:Acc} we study the accuracy we can reach in the determination of $\nu_{\rm max}$ for different observational times and duty cycles. Section \ref{sec:Prec} is devoted to the analysis of the stellar parameters precision we can obtain when the observable $\nu_{\rm max}$ is observed with the accuracy estimated in the previous section. Finally, section \ref{sec:concl} is devoted to summarize the conclusions of this study. In the Appendix \ref{sec:scuti} we analyze the potential of {\it CHEOPS} for observing $\delta$~Scuti's $\nu_{\rm max}$ and the accuracy we can reach.

\section{Estimation of the detectability potential of solar-like oscillations using {\it CHEOPS}}
\label{sec:est_gen}

Solar-like pulsations are reflected in the frequency power spectrum as a group of peaks with power amplitudes following a Gaussian-like profile. Following \citet{Chaplin11} and \citet{Campante16}, the estimation of the detectability potential of solar-like pulsations is, therefore, obtained by analyzing in the frequency power spectrum the probability of a Gaussian-like power excess to be statistically different from noise.

This analysis is done in three steps: First, using semi-empirical relations and stellar models, we estimate the expected strength in the power spectrum of the stellar pulsations ($P_{\rm tot}$). Second, we estimate the background power density coming from different sources ($B_{\rm tot}$). Finally, we evaluate the probability of the estimated power amplitudes to be statistically different from noise, using the expected signal to noise ratio ($S/N_{\rm tot}=P_{\rm tot}/B_{\rm tot}$).

\subsection{Expected power spectrum amplitudes, $P_{\rm tot}$}
\label{sec:Ptot}

The expected contribution of the pulsational modes to the power spectrum may be approximated by

\begin{equation}
P_{\rm tot} \approx 0.5cA^2_{\rm max}\eta^2(\nu_{\rm max})D^{-2}\frac{W}{\Delta \nu} \;\;\;\mathrm{ppm}^2
\end{equation}

\noindent where $A_{\rm max}$ is the expected maximum amplitude of the radial modes ($l=0$) in parts per million (ppm). The factor $c$ measures the mean number of modes per $\Delta \nu$ segment and depends on the observed wavelength. Since {\it CHEOPS} bandpass is similar to that of {\it Kepler} \citep{Gaidos17}, we have used the same $c$ calculated by \citet{Chaplin11} following \citet{Bedding96} ($c=3.1$). The attenuation factor $\eta^2(\nu)=\mathrm{sinc}^2\Big[\frac{\pi}{2}\frac{\nu}{\nu_{\rm Nyq}}\Big]$ takes into account the apodization of the oscillation signal due to the non-zero integration time in the case of an integration duty cycle of $100\%$, where $\nu_{\rm Nyq}$ is the Nyqvist frequency. $\nu_{\rm max}$ is the frequency of the maximum power spectrum in the stellar pulsation regime. $D$ is a dilution factor defined as $D=1$ for an isolated object. In this study we will focus in this isolated star case. $\Delta \nu$ is the so-called large separation or the distance between neighboring overtones with the same spherical degree $l$. On average, the power of each $\nu_{\rm max}/\Delta\nu$ segment will be $ \sim$0.5 times that of the central segment, thus explaining the extra 0.5 factor. Finally, $W$ is the range where the power of the pulsational mode is contained. \citet{Mosser12, Mosser10} and \citet{Stello07} estimated that

\begin{equation}
    W(\nu_{\rm max})= 
\begin{cases}
    1.32\nu^{0.88}_{\rm max},& \text{if } \nu_{\rm max} \leq 100 \mu {\rm Hz}\\
    \nu_{\rm max},& \text{if } \nu_{\rm max} > 100 \mu {\rm Hz}
\end{cases}
\label{eq:W}
\end{equation}

The expected $A_{\rm max}$ can be estimated using stellar models from the following semi-empirical relation

\begin{equation}
A_{\rm max}=2.5\beta\left(\frac{R}{R_\odot}\right)^2\left(\frac{T_{\rm eff}}{T_{\rm eff,\odot}}\right)^{0.5}\;\;\;\mathrm{ppm}
\end{equation}

This estimation was first derived by \citet{Chaplin11} for the case of the observations of {\it Kepler} and it depends on stellar parameters and the instrument response filter. In our case, we can use the same expression without any correction, where $R$ and $T_{\rm eff}$ are the stellar radius and effective temperature, respectively, and $R_\odot$ and $T_{\rm eff,\odot}$ the solar values. On the other hand, $\beta$ is a factor introduced to correct the overestimation that this expression does of the amplitudes for the hottest stars. That is,

\begin{equation}
\beta = 1-exp\left(-\frac{T_{\rm red}-T_{\rm eff}}{1550\,{\rm K}}\right)
\end{equation}

\noindent where $T_{\rm red}$ is the blue boundary of solar-like oscillations (or the red boundary of $\delta$~Scuti pulsations) and its empirical estimation is

\begin{equation}
T_{\rm red}=8907\left(\frac{L}{L_{\odot}}\right)^{-0.093}\;\;\;\mathrm{K}
\end{equation}

\noindent where $L$ is the stellar luminosity and $L_\odot$ is the corresponding solar value.

For a detailed explanation of the origin and assumptions of these expressions, we refer to the papers of \citet{Chaplin11} and \citet{Campante16}.

Although these estimations have been done under the assumption of an integration duty cycle of 100$\%$, duty cycles larger than 60$\%$, as it is the case of CHEOPS light-curves, have a negligible impact in the estimated power excess and in the obtaining of $\nu_{\rm max}$ \citep{Stahn08}

\subsection{Estimation of the total background power, $B_{\rm tot}$}
\label{sec:Btot}

The background power spectral density in the zone of $\nu_{\rm max}$ can be approximated as

\begin{equation}
B_{\rm tot}\approx b_{\rm max}\,W(\nu_{\rm max})\;\;\;\mathrm{ppm}^2
\end{equation}

The main contributions to $b_{\rm max}$ are assumed to be the instrumental/astronomical noises (jitter, flat field, timing error, etc., on the one hand, photon noise, zodiacal light, etc., on the other), and the stellar granulation. This second contribution has a significant impact when the observations of the oscillations are made using photometry. That is:

\begin{equation}
b_{\rm max}=b_{\rm instr} + P_{\rm gran}\;\;\;\mathrm{ppm}^2\mu {\rm Hz}^{-1}
\end{equation}

\subsubsection{Instrumental/astronomical noise, $b_{\rm instr}$}

Following \citet{Chaplin11},

\begin{equation}
b_{\rm instr}=2\times 10^{-6}\sigma^2 \Delta t\;\;\;\mathrm{ppm}^2\mu {\rm Hz}^{-1}
\label{eq:instr_sigma}
\end{equation}

\noindent where $\sigma$ is the {\it CHEOPS} predicted {\it RMS} noise per a given exposure time ($T_{\rm exp}$) and $\Delta t$ the integration time. Following the CHEOPS Red Book noise budget \citep{CHEOPS_RED}, the {\it RMS} noise is the addition of several contributions, but the final value mainly depends on the stellar magnitude, the exposure time (linked to the stellar magnitude), and the integration time. Regardless of the exposure time, {\it CHEOPS} adds and downloads images every 60 seconds. This will be our integration time. Taking a look to Equation \ref{eq:instr_sigma}, we see that $b_{\rm instr}$ is almost independent of $\Delta t$, since $\sigma^2 \sim 1/\Delta t$. Therefore, we will work only with one integration time: 60 sec. In Table \ref{tab:sigma} we show the instrumental {\it RMS} for three different stellar magnitudes ($V$=6, 9, and 12), with three corresponding exposure times (1, 10, and 60 sec., respectively). These values are our reference values.

\begin{table}
	\centering
	\caption{Different instrument/astronomical noises of {\it CHEOPS} ($\Delta t=60$ s).}
	\begin{tabular}{ccc}
		\hline
		Magnitude   & $T_{\rm exp}$  & $\sigma_{\rm Nom}$  \\
		(in $V$) & (in s) & (in ppm)\\
		\hline
		6   & 1 & 44\\
		9  &  10 &  165\\
		12 & 60 & 623\\
		\hline
	\end{tabular}
	\label{tab:sigma}
\end{table}

\subsubsection{Granulation power spectrum density, $P_{\rm gran}$}

Following \citet{Campante16}, we use the model F of \citet{Kallinger14} (with no mass dependence). Evaluating this model at $\nu_{max}$ we have

\begin{equation}
P_{\rm gran,real}(\nu_{\rm max})=\eta^2(\nu_{\rm max})D^{-2}\sum_{i=1}^2\frac{\left(\frac{2\sqrt{2}}{\pi}\right)\frac{a_i^2}{b_i}}{1+\left(\frac{\nu_{\rm max}}{b_i}\right)^4}\;\;\;\mathrm{ppm}^2\mu {\rm Hz}^{-1}
\end{equation}

\noindent where

\begin{subequations}
\begin{eqnarray}
a_{1,2}  & = & 3382\,\nu_{\rm max}^{-0.609}\\
b_1 & = & 0.317\,\nu_{\rm max}^{0.970}\\
b_2 & = & 0.948\,\nu_{\rm max}^{0.992}
\end{eqnarray}
\end{subequations}

The parameters of these models have been fitted using the power spectra of a large set of {\it Kepler} targets. Therefore, we can use these estimations without any correction for {\it CHEOPS} thanks to their similar response filters.

Another effect to take into account when modeling granulation is aliasing. When the observed signal has frequencies above the Nyqvist frequency, they can appear in the power spectrum at sub-Nyqvist frequencies since these frequencies are undersampled, contributing to the background noise. With a cadence of 60 sec and its associated large $\nu_{\rm Nyq}$, the impact of this aliased granulation power in the final background noise is small, but we have included it for completeness. Following \citet{Campante16}, the aliased granulation power at $\nu_{\rm max}$ can be modeled as $P_{\rm gran,aliased}(\nu_{\rm max})\equiv P_{\rm gran,real}(\nu_{\rm max}^\prime)$, where

\begin{equation}
    \nu_{\rm max}^\prime= 
\begin{cases}
    \nu_{\rm Nyq}+(\nu_{\rm Nyq}-\nu_{\rm max}),& \text{if } \nu_{\rm max} \leq \nu_{\rm Nyq}\nonumber\\
    \nu_{\rm Nyq}-(\nu_{\rm max}-\nu_{\rm Nyq}),& \text{if } \nu_{\rm Nyq} < \nu_{\rm max} \leq 2\nu_{\rm Nyq}
\end{cases}
\end{equation}

\noindent and then $P_{\rm gran} = P_{\rm gran,real}(\nu_{\rm max})+P_{\rm gran,aliased}(\nu_{\rm max})$.

\subsection{Asteroseismic scaling relations}

In Sections \ref{sec:Ptot} and \ref{sec:Btot} we have seen that for the estimation of the expected spectrum power excess and background, the asteroseismic parameters $\Delta \nu$ and $\nu_{\rm max}$ must be known. The most efficient way of doing so is by means of the so-called scaling relations. These relations explode the fact that $\Delta \nu$ and $\nu_{\rm max}$ are a function of the stellar mean density and surface gravity respectively. Therefore, they can be approximately estimated when stellar mass, radius, and effective temperature have reasonable estimates. In our study these stellar parameters, using solar metallicity, are provided by stellar models obtained using the evolutionary code CLES \citep[Code Li\'egeois d'Evolution Stellaire,][]{2008Ap&SS.316...83S}. Then, using the scaling relations shown in \citet[]{Kallinger10_2}, and references therein, we estimate $\Delta \nu$ and $\nu_{\rm max}$. That is:

\begin{eqnarray}
\nu_{\rm max} & \approx & \nu_{\rm max,\odot}\left(\frac{M}{M_{\odot}}\right)
\left(\frac{R}{R_{\odot}}\right)^{-2}
\left(\frac{T_{\rm eff}}{T_{\rm eff,\odot}}\right)^{-0.5}\label{eq:numax}\\
\Delta \nu & \approx & \Delta \nu_{\odot}\left(\frac{M}{M_{\odot}}\right)^{0.5}
\left(\frac{R}{R_{\odot}}\right)^{-1.5}
\label{eq:deltanu}
\end{eqnarray}

\noindent where the solar reference values are $\nu_{\rm max,\odot}=3090\mu {\rm Hz}$ and $\Delta \nu_{\odot}=135.1\mu {\rm Hz}$.

\subsection{Estimation of the detection probability}

From stellar models, together with some instrument prescriptions, we can estimate the expected signal to noise ration as $S/N_{\rm tot}=P_{\rm tot}/B_{\rm tot}$, where $P_{\rm tot}$ and $B_{\rm tot}$ are known.

Estimating the detection probability, in this context, is to test whether a given $S/N_{\rm tot}$ can be randomly produced from noise or whether it is a signal of a statistic significant power excess in the original data.

As we mentioned in the introduction of this section, the solar-like pulsational power excess in the frequency spectrum has a Gaussian-like profile. On the other hand, if $T$ is the length of the time series, the information in the frequency domain comes in bins of $1/T ({\rm s}^{-1})$, and the number of bins contained in the potential zone of solar-like pulsation power excess is $N=W \times T$. The $N$ we use in this work has been obtained assuming a duty cycle of a 100$\%$. Nevertheless, following \citet{App14, Campante12}, the impact of duty cycles in the range [60 - 100]$\%$, as it is the case of {\it CHEOPS}, is small. Therefore, the statistical test we must perform is to disentangle whether a Gaussian-like profile described using $N$ bins can be produced by a random distribution or not.

This problem is faced using a $\chi^2$ test with $2N$ degrees of freedom following \citet{Appourchaux04}. The first step in this test is to fix a minimum threshold avoiding a false alarm positive ($S/N_{thres}$). This limit is fixed at a 5$\%$ of a chance of false positive (p-value < 0.05). That is, if we define $x=1+S/N$, the p-value of this test is

\begin{equation}
p=\int_x^{\infty}\frac{exp(-x^\prime)}{\Gamma(N)}x^{\prime (N-1)}dx^\prime
\end{equation}

\noindent where $\Gamma$ is the Gamma function. $S/N_{\rm thres}$ is the one making $p=0.05$. Since $N$ is a function of $\nu_{\rm max}$, every model has its own threshold.

Once $S/N_{\rm thres}$ is obtained, the probability of a given excess to be statistically different from random noise is

\begin{equation}
p_{\rm excess} = \int_y^{\infty}\frac{exp(-y^\prime)}{\Gamma(N)}y^{\prime (N-1)}dy^\prime
\end{equation}

\noindent where $y=(1+S/N_{\rm thres})/(1+S/N_{\rm tot})$.

Therefore, for a given stellar model ($M$, $R$, $T_{\rm eff}$, $\Delta \nu$, and $\nu_{\rm max}$), an observational time range ($T$), and a stellar magnitude, we can calculate the probability of the solar-like pulsations to be observed.

\begin{figure}
 \includegraphics[width=\columnwidth]{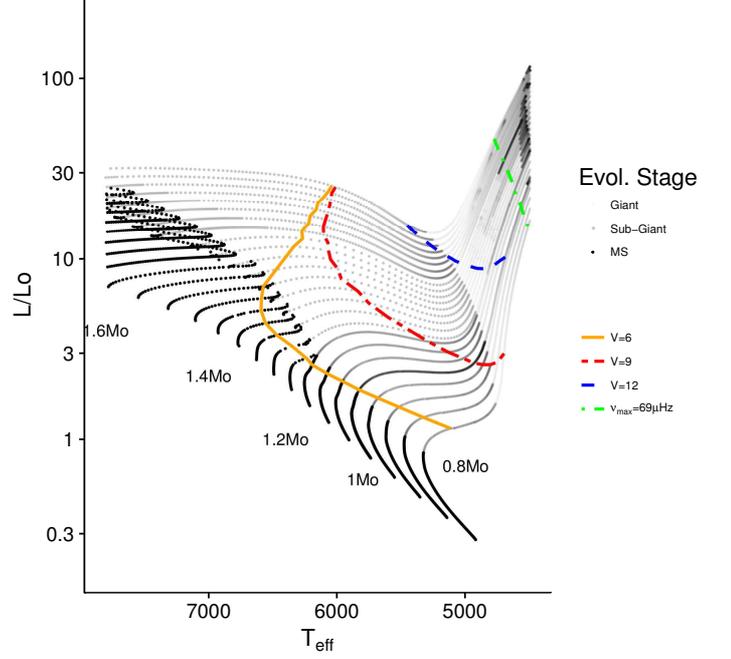}
 \caption{HR Diagram showing the region where $\nu_{\rm max}$ can be determined using eight hours of {\it CHEOPS} observations. Black points are the models used aligned in different evolutionary tracks. The evolutionary stage of these models are represented in different point transparency: MS stars in black, Sub-Giant dark gray, and Giant in light gray. The orange line is the bottom limit for a potential detection of a star with Magnitude $V$=6. The same bottom limit for the case of a star with $V$=9 is shown with a red line. The blue line is the bottom limit for a $V$=12 star. The green line represents the top limit for a correct coverage of the variability with eight hours of observational time.}
 \label{fig:HR_prob_1}
\end{figure}

In Fig. \ref{fig:HR_prob_1} we show a HR diagram with the evolutionary tracks used for the simulations and the bottom limit for a $p_{\rm excess} =50\%$ of detection probability for three different stellar magnitudes (orange, red and blue lines for
$V =$ 6, 9, and 12 stars respectively). Every star above these lines has a $\nu_{\rm max}$ potentially detectable by {\it CHEOPS}. On the other hand, integration and total observing times may put some constraints to our observational capabilities:

\begin{figure}
 \includegraphics[width=\columnwidth]{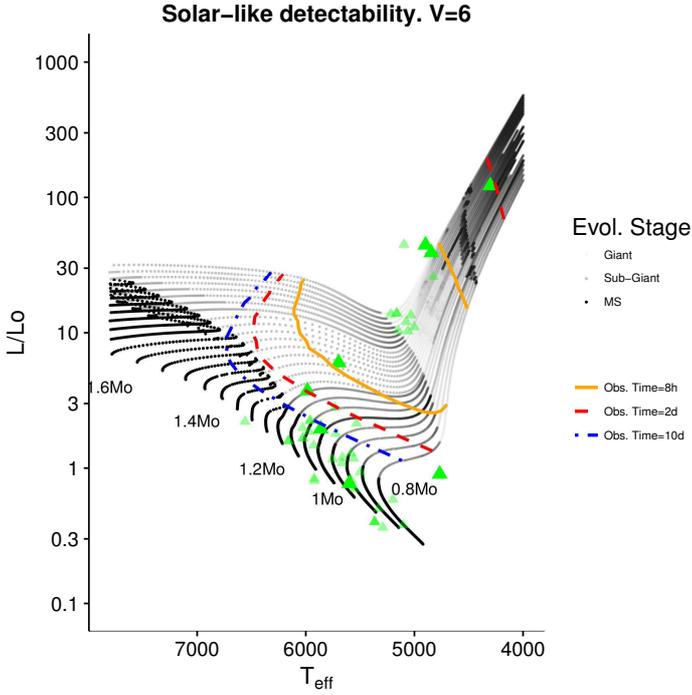}
 \caption{HR Diagram showing the region where $\nu_{\rm max}$ can be determined, for a star with magnitude $V
 $=6, using different observational times. Black points are the models used aligned in different evolutionary tracks. The evolutionary stage of these models are represented in different point transparency: MS stars in black, Sub-Giant dark gray, and Giant in light gray. The orange lines are the bottom and top limits of the potentially $\nu_{\rm max}$ detection region for eight hours of observational time (respectively red lines - two days of observational time, blue lines - ten days of observational time). The upper detectability limit for an observational time of 10 days is outside the plotted range. The position of the current known stars with planets in the ecliptic plane with $V$<6 is shown using green triangles. The small and more transparent green triangles are those known stars with planets in the ecliptic plane with 6<$V$<9.}
 \label{fig:HR_prob_1_obs_range}
\end{figure}

\begin{figure}
 \includegraphics[width=\columnwidth]{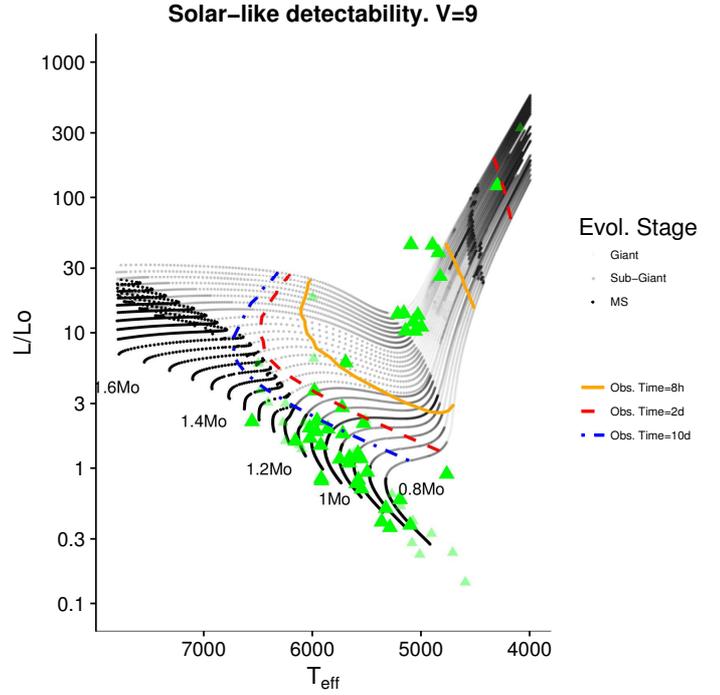}
 \caption{Same as Fig. \ref{fig:HR_prob_1_obs_range} for a star with magnitude $V$=9. The position of the current known stars with planets in the ecliptic plane with $V$<9 is shown using green triangles. The small and more transparent green triangles are those known stars with planets in the ecliptic plane with 9<$V$<12.}
 \label{fig:HR_prob_1_obs_range_9M}
\end{figure}

\begin{figure}
 \includegraphics[width=\columnwidth]{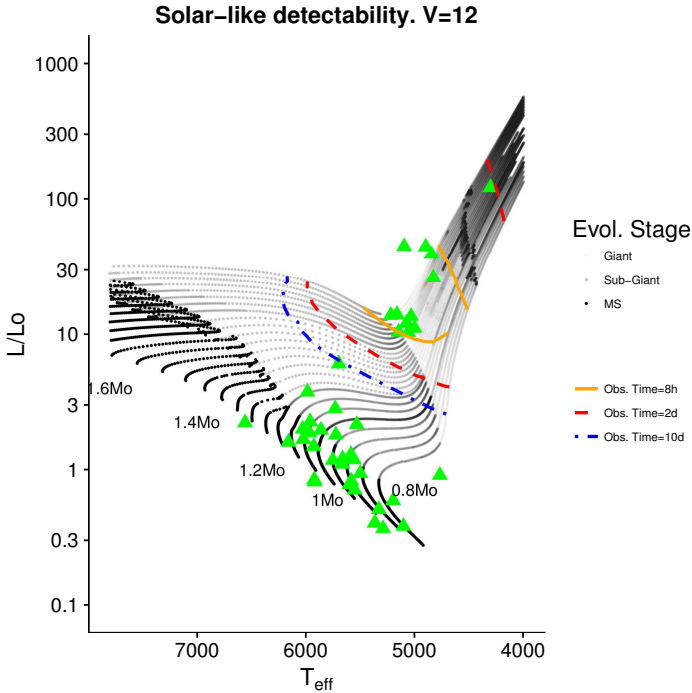}
 \caption{Same as Fig. \ref{fig:HR_prob_1_obs_range} for a star with magnitude $V$=12. The position of the current known stars with planets in the ecliptic plane with $V$<12 is shown using green triangles.}
 
 \label{fig:HR_prob_1_obs_range_12M}
\end{figure}
\begin{itemize}
\item {\it Integration time}: As we have already explained, changing the integration time is not efficient in our context and we have fixed it to 60 sec, imposing limits to the $\nu_{\rm max}$ that can be properly monitored. Each stellar model has its own $\nu_{max}$, depending on stellar parameters following Eq. \ref{eq:numax}. The largest $\nu_{\rm max}$ is located at the orange line of Fig. \ref{fig:HR_prob_1} with a value of $1765\,\mu Hz$, that is, around 9.5 min. Therefore, the integration time of 1 min is enough for covering properly every potentially detectable $\nu_{\rm max}$ ($\nu_{\rm Nyq}$=8333 $\mu$HZ, covering all the frequency range).

\item {\it Total observing time}: One of the most conservative goals of our time series is to monitor at least one complete period for all the solar-like modes at the expected frequency range. For a given $\nu_{\rm max}$, eq. \ref{eq:W} shows that the shortest expected frequency (largest expected period) is, roughly, 0.5$\nu_{\rm max}$. That is, for a given total observational time ($T$), we can ensure to monitor at least one complete period for all the expected solar-like modes of stars with $\nu_{\rm max}= 2/T$. The green line in Fig. \ref{fig:HR_prob_1} represent the limit of $\nu_{\rm max} = 69\,\mu {\rm Hz}$. Every star below this limit can be correctly monitored with eight hours of observing time.
\end{itemize}

Therefore, with the most conservative election of the different degrees of freedom, every star with characteristics in the region limited by the orange/red/blue lines and the green line of Fig. \ref{fig:HR_prob_1} has a $\nu_{\rm max}$ potentially detectable with {\it CHEOPS} with eight hours of observational time.


In any case, the total observing time can be modified. In Figs. \ref{fig:HR_prob_1_obs_range} to \ref{fig:HR_prob_1_obs_range_12M} we show the impact of changing the observational time up to ten days in the definition of the potentially detectable region.

Although the general observational strategy of {\it CHEOPS} is to spend several hours per target, in some special cases a larger observational time will be accessible, as it is the case of orbiting phase studies. In Fig. \ref{fig:HR_prob_1_obs_range} we show the impact of increasing this observational time. The three lines represent the theoretical limits, for a 6th magnitude star, as a function of the total observational time. This observational time impacts in two ways: on one hand, increasing the accuracy of the signal mapping in the frequency domain (reducing the size of the bins), and on the other hand, allowing the study of larger periods. That is, it pushes down the bottom limit for potential detectability and pushes up the top limit in the Red Giant Branch. The impact of having eight hours of observing time (orange lines), two days (red lines), or ten days (blue lines) is what we show in Figs. \ref{fig:HR_prob_1_obs_range}, \ref{fig:HR_prob_1_obs_range_9M}, and \ref{fig:HR_prob_1_obs_range_12M} for a $V=$ 6, 9, and 12 star respectively. The increasing of the area enabling a potential $\nu_{\rm max}$ characterization is remarkable.

\begin{table*}
	\centering
	\caption{Summary of the sampling of stars with planets in the region Dec. = [-6,6]$^\circ$.}
	\begin{tabular}{ccc}
		\hline
		Charact. & Min & Max \\
		\hline
		$T_{\rm eff}$ ($^\circ$K)&940 &33000 \\
		log$L_{\rm bol}$ (dex)& -6.5& 2.5\\
		$[$Fe/H$]$ (dex)& -0.71 & 0.56\\
		\hline
		Detecting & Number & \\
		Method & & \\
		\hline
		Imaging & 5 & \\
		Transit & 97 & \\
		Pulsar & 3 & \\
		RV & 64 & \\
		\hline
	\end{tabular}
	\label{tab:exopl_sampling}
	\tablebib{Taken from www.exoplanet.org \citep{exopl}. $L_{\rm bol}$ have been obtained using VOSA \citep{vosa}.
	}
\end{table*}

In the Introduction we mentioned that {\it CHEOPS} and {\it TESS} are complementary in terms of their asteroseismic capabilities. We have identified that, nowadays, there are 169 stars harbouring planets located at the ecliptic plane (Dec. = [-6,6]$^\circ$). In table \ref{tab:exopl_sampling} we show a summary of the main characteristics of this sampling. In Figs. \ref{fig:HR_prob_1_obs_range}, \ref{fig:HR_prob_1_obs_range_9M}, and \ref{fig:HR_prob_1_obs_range_12M} we have also shown the position in the HR-Diagram of a subsample of these 169 stars, those within the magnitude constraints of each figure and their $T_{\rm eff}$ and $L/L_\odot$ limits. We have shown in big green triangles those stars of this sample brighter than the limit displayed at the plot. In smaller and more transparent green triangles we show those stars with apparent magnitudes between that presented in the plot and that of the following plot. That is, in Fig. \ref{fig:HR_prob_1_obs_range} the big triangles are stars with $V$<6, and the small triangles are stars with 6<$V$<9.

Therefore, as a summary, we can conclude that {\it CHEOPS} can potentially observe the solar-like pulsation characteristic $\nu_{\rm max}$ for FGK stars using its planned standard observational strategy in the following cases: Massive stars ($1.4M_\odot < M < 1.2M_\odot$) from late MS on, and post-MS for all the masses. Depending on the observational time, tens of stars with planets located at the ecliptic plane can be characterized with a larger precision (see sections \ref{sec:Acc} and \ref{sec:Prec}).

Increasing the observational time within {\it CHEOPS} accepted observational strategy has a large impact on the HR diagram zone of potential detectability. The larger the observational time, the larger the MS region potentially covered.

\section{Impact of observational time and duty cycle in the accuracy of the determination of $\nu_{\rm max}$}
\label{sec:Acc}

In the previous section, we have analyzed whether solar-like pulsations can be potentially detectable with {\it CHEOPS} depending on the different instrumental and observational constraints, but we have not described which observables can be obtained from these observations and the accuracy of this characterization.

As we have already mentioned, the total observational time per target will be of the order of hours or a few days. This is not enough for disentangling individual frequencies (discarding $\Delta \nu$ as a possible observable), but it is enough for obtaining the frequency with the maximum power amplitude. Therefore, we will focus our studies on how accurately we can determine $\nu_{\rm max}$ and the impact of this additional observable in the characterization of the targets.

\begin{table*}
	\centering
	\caption{Testing stars.}
	\begin{tabular}{ccccccccccc}
		\hline
		KIC & $T_{\rm eff}$ & $\Delta T_{\rm eff}$ & ${\rm log}g$ & $\Delta {\rm log}g$ & [Fe/H] & $\Delta$[Fe/H] & $\nu_{\rm max}$ & $\Delta\nu_{\rm max}$ & Reference & S/N\\
		&  in K& & in dex& & in dex& & in $\mu$Hz & & &\\
		\hline
		1435467 &6326 & 77 & 4.100 & 0.1 & 0.01 & 0.1 & 1406.7 & 8.4 & a & 60\\
		3456181 &6384 & 77 & 3.950 & 0.1 & -0.15 & 0.1 & 970.0 & 8.3 & a & 60\\
		5701829 &4920 & 100 & 3.19 & 0.22 & -0.2 & 0.2 & 148.3 & 2.0 & b & 200\\
		6933899 &5832 & 77 & 4.079 & 0.1 & -0.01 & 0.1 & 1389.9 & 3.9 & a &200\\
		7771282 &6248 & 77 & 4.112 & 0.1 & -0.02 &  0.1 & 1465.1 & 27.0 & a & 70\\
		9145955 &4925 & 91 & 3.04 & 0.11 &  -0.32 & 0.03 & 131.7 & 0.2 & c & 200\\
		9414417 &6253 & 77 & 4.016 & 0.1 & -0.13 & 0.1 & 1155.3 & 6.1 & a & 80 \\
		9812850 &6321 & 77 & 4.053 & 0.1 & -0.07 &  0.1 & 1255.2 & 9.1 & a & 60\\
		10162436 &6146 & 77 & 3.981 & 0.1 & -0.16 & 0.1 & 1052.0 & 4.2 & a & 100\\
		12069127 &6276 & 77 & 3.912 & 0.1 &  0.08 &  0.1 & 884.7 & 10.1 & a& 60\\
		\hline
	\end{tabular}
	\label{tab:star_test}
	\tablebib{(a)~\citet{Lund17};
		(b) \citet{Fox-Machado16}; (c) \citet{Perez16}.
	}
\end{table*}

\begin{table}
	\centering
	\caption{Tested duty cycles.}
	\begin{tabular}{ccc}
		\hline
		Simulation & \multicolumn{2}{c}{Duty Cycle ($\%$)}\\
		& SAA $\&$ EO & Complete  \\
		\hline
		1 &100.0& 90.4\\
		2 &90.6 &86.7\\
		3 &80.4 &79.9\\
		4 &72.5 &72.1\\
		5 &72.1 &71.7\\
		6 &68.0 &67.6\\
		7 &65.0 &64.6\\
		8 &62.9 &62.6\\
		9 &61.5 &61.1\\
		10 &60.4 &60.0\\
		11 &59.5 &59.2\\
		\hline
	\end{tabular}
	\label{tab:duty_cycle}
\end{table}

\begin{figure*}
\subfloat[]{\includegraphics[width=0.66\columnwidth]{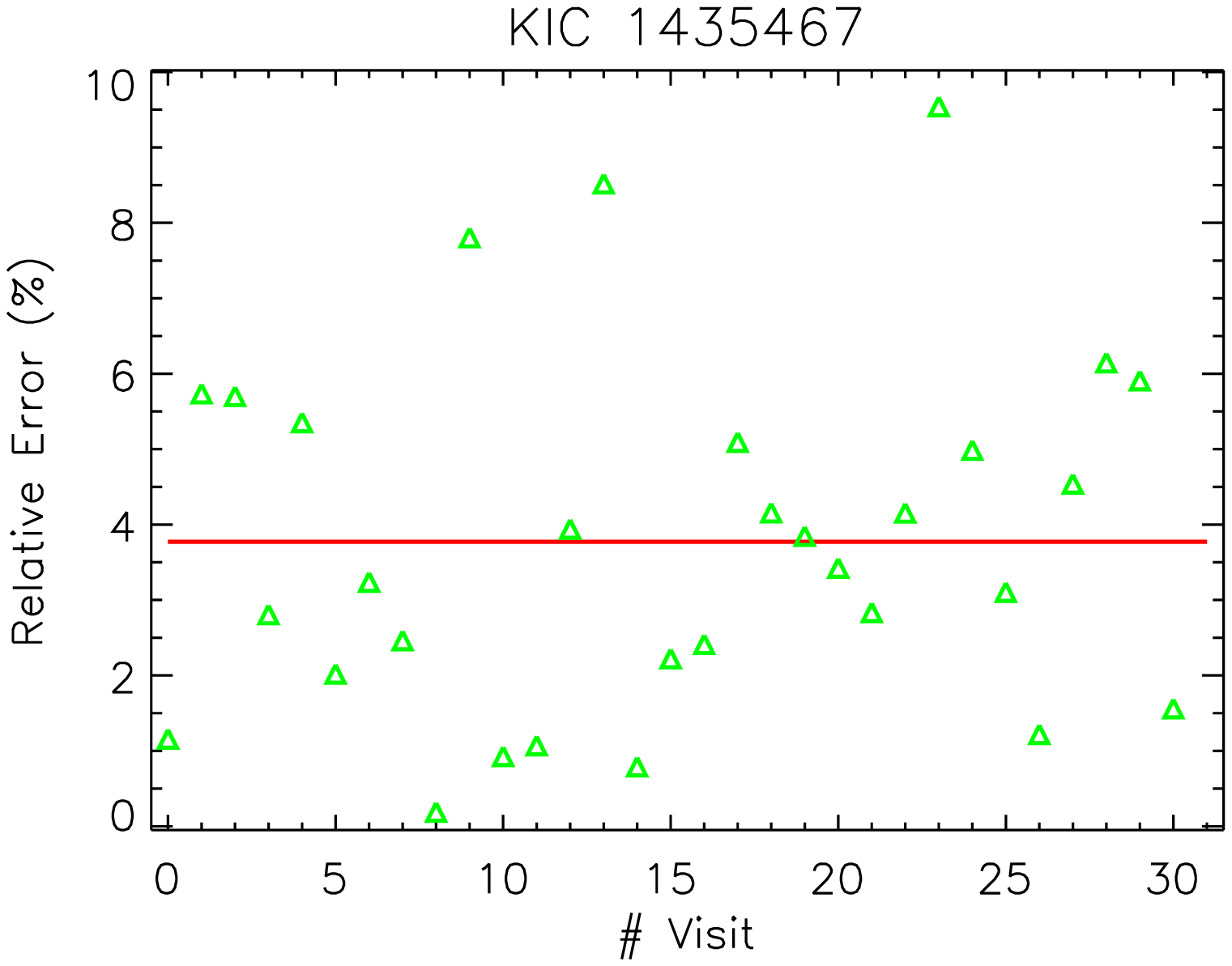}}
\subfloat[]{\includegraphics[width=0.66\columnwidth]{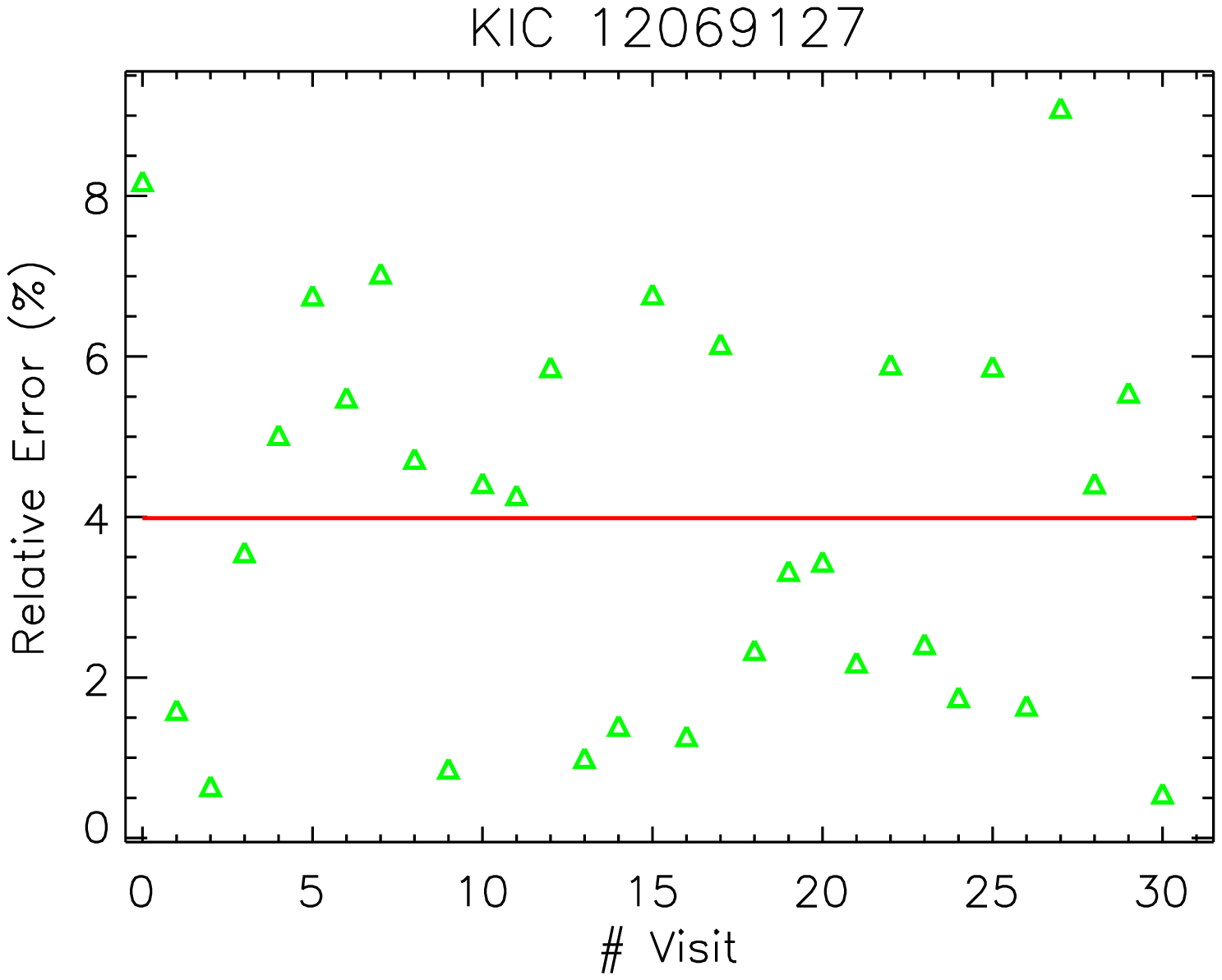}}
\subfloat[]{\includegraphics[width=0.66\columnwidth]{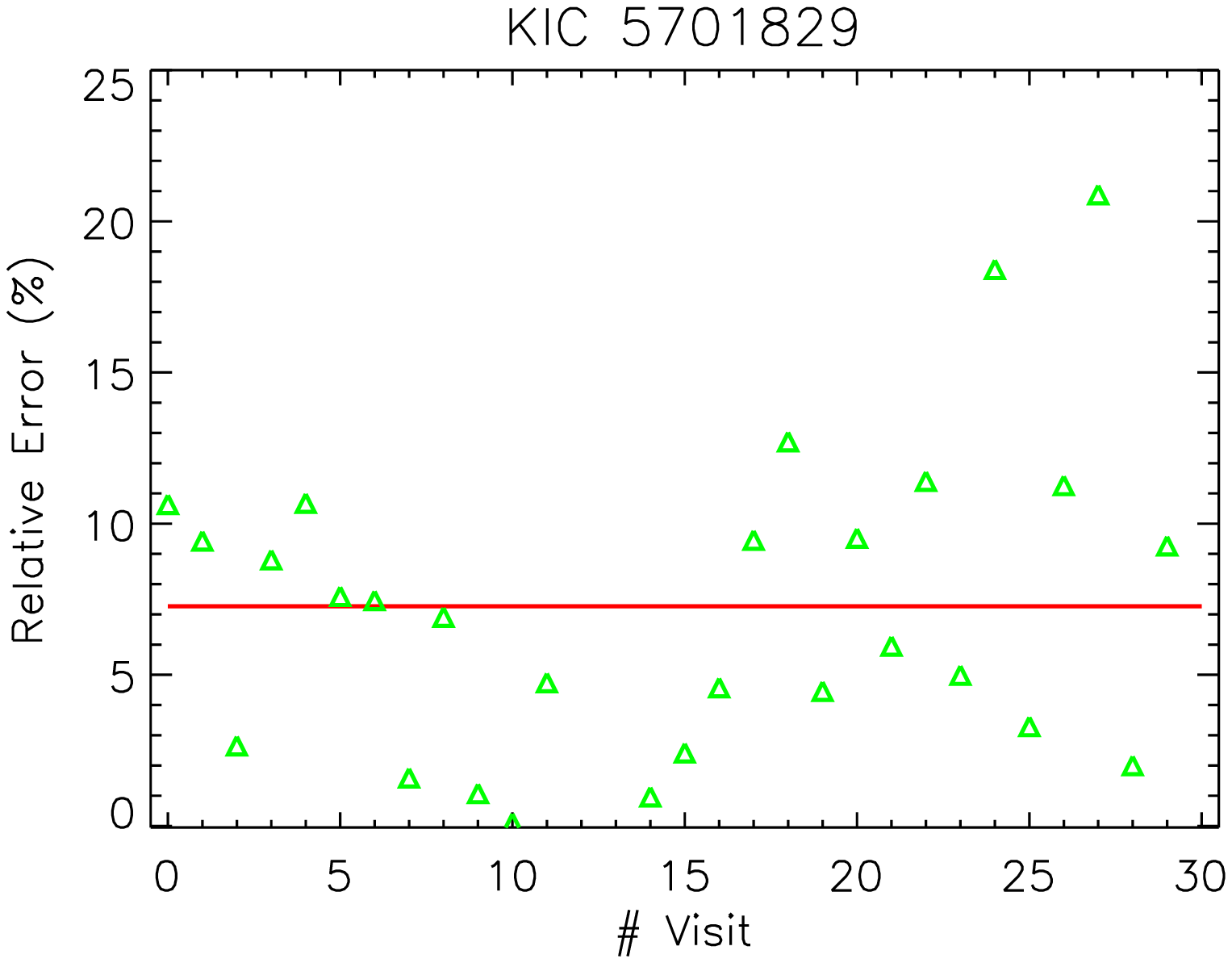}}
 \caption{Relative error of $\nu_{\rm max,w}$ calculated with an 1-day run, a $80\%$ duty cycle, and taking into account the case 1. Only those significant and inside the detection range values are shown. From left to right, we show the results of a Main Sequence star (a), a subgiant (b), and a RGB star (c). The black/green asterisks points to all non-/significant} points inside the detection range (purple line). The red line is the relative error of the mean $\nu_{\rm max,w}$.
 \label{fig:dc_1}
\end{figure*}

\begin{figure*}
\resizebox{\hsize}{!}{\includegraphics[width=\columnwidth]{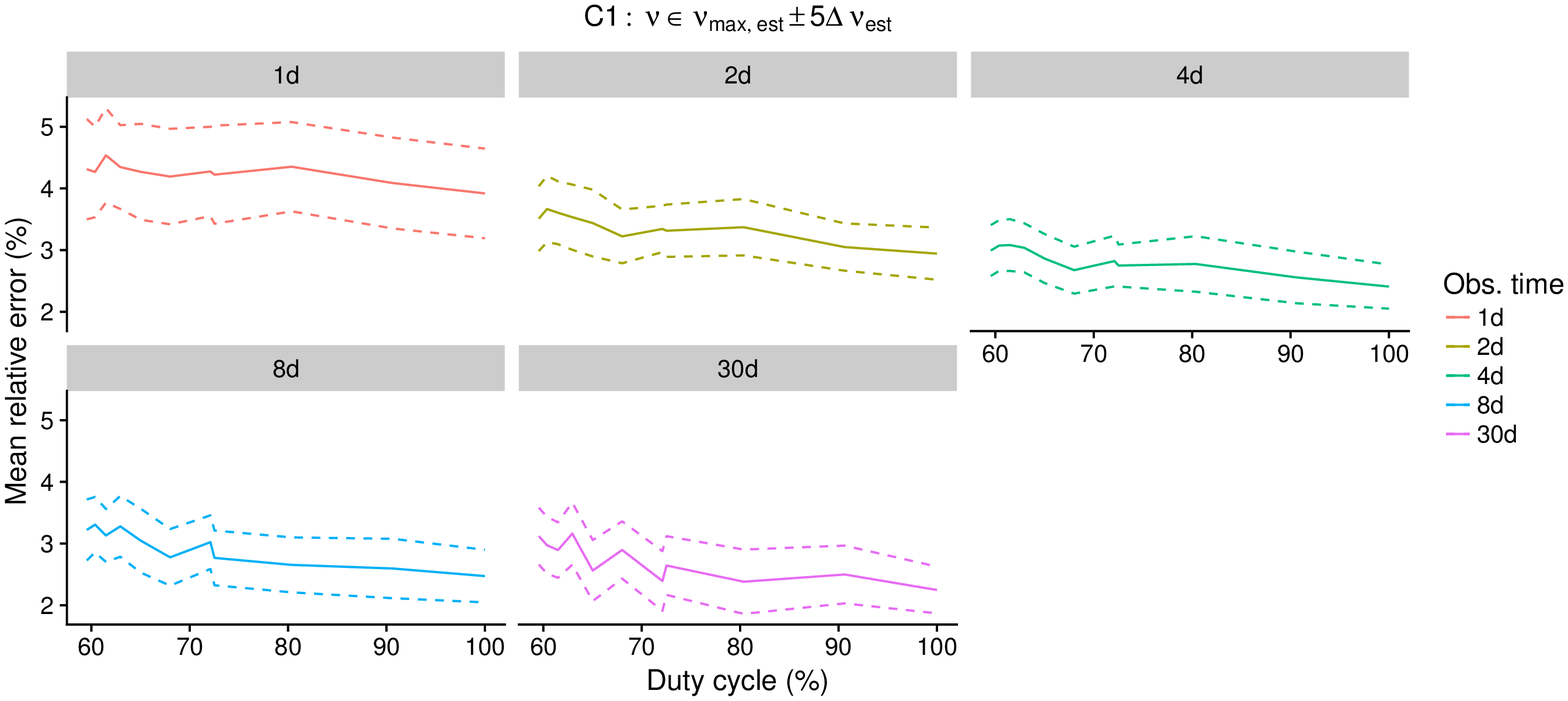}}
 \caption{Mean relative error within the range $\pm 5\Delta \nu_{est}$ (C1) according to the duty cycle for different observational times.  Dashed lines are $\pm$ their standard deviation}
 \label{fig:mean_1}
\end{figure*}

\begin{figure*}
\resizebox{\hsize}{!}{\includegraphics[width=\columnwidth]{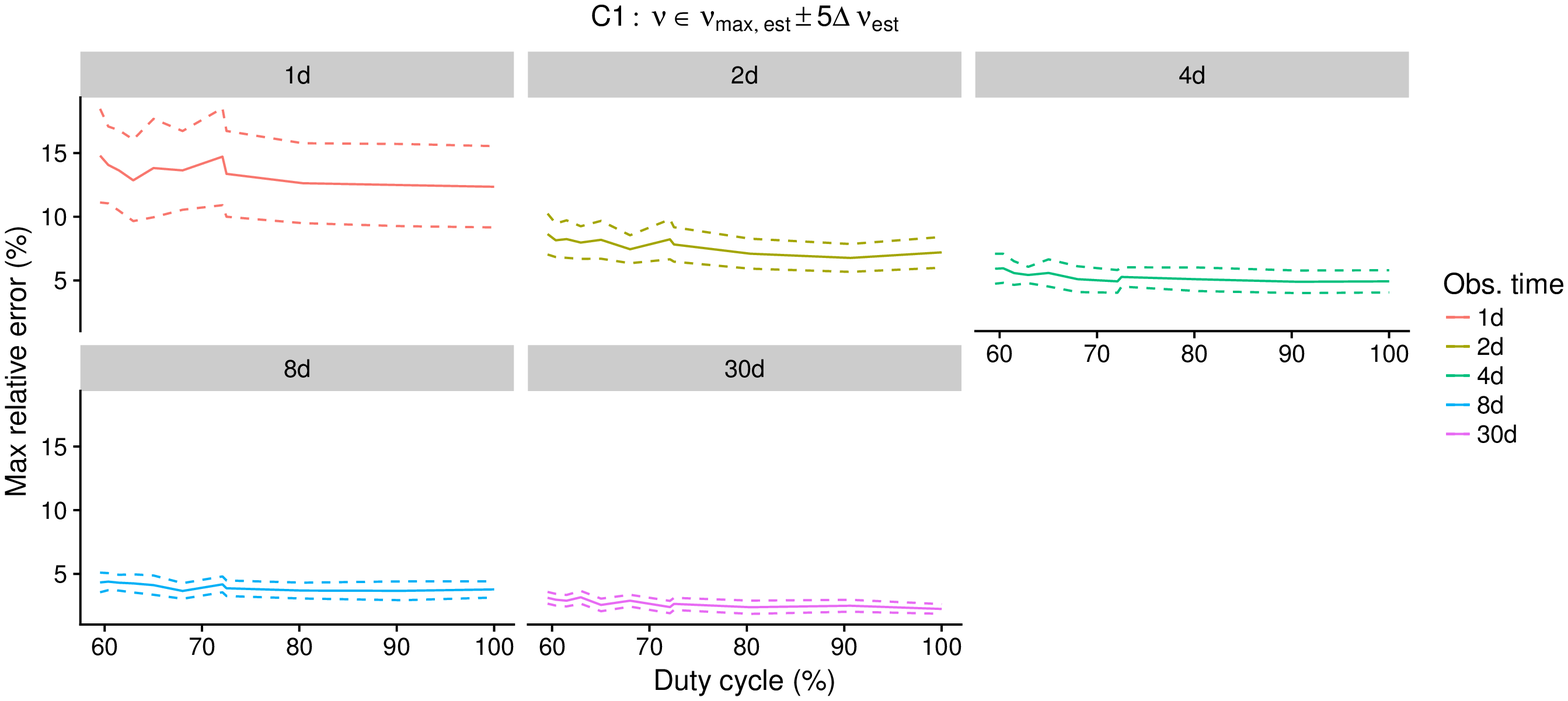}}
 \caption{Maximum relative error within the range $\pm 5\Delta \nu_{est}$ (C1) according to the duty cycle for different observational times. Dashed lines are $\pm$ their standard deviation}
 \label{fig:max_1}
\end{figure*}

\begin{figure*}
\resizebox{\hsize}{!}{\includegraphics[width=\columnwidth]{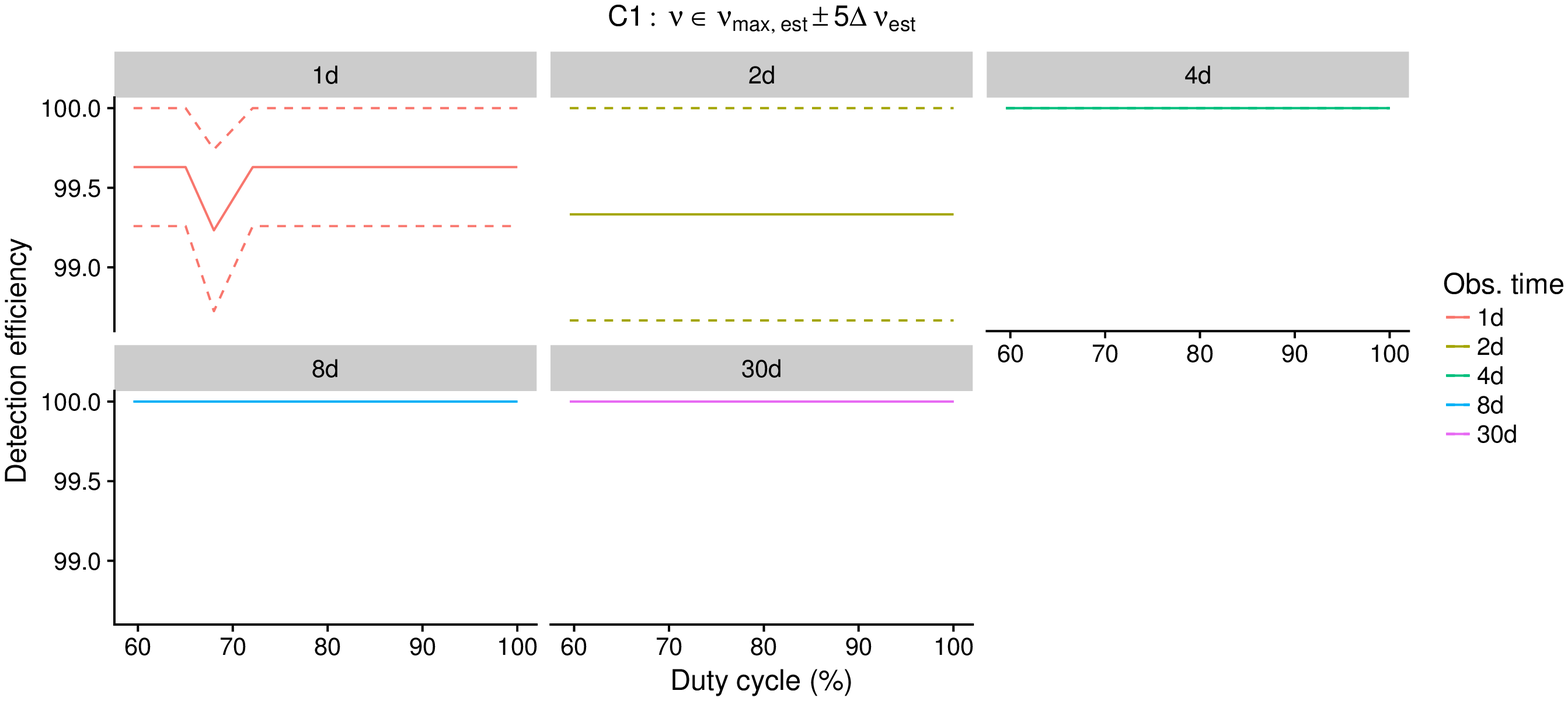}}
 \caption{Efficiency of detection within the range $\pm 5\Delta \nu_{est}$ (C1) according to the duty cycle for different observational times. Dashed lines are $\pm$ their standard deviation}
 \label{fig:prob_1}
\end{figure*}

The total observational time ($T$) has an impact in the HR diagram zone where $\nu_{\rm max}$ can be potentially detected, but it has also an impact on the accuracy of the determination of its value from observations. $T$ has a major impact on the definition of the numbers of bins we will have in the frequency domain ($N=W \times T$). The larger the $T$, the larger the number of bins and, therefore, the better the mapping of this frequency space. Since $\nu_{\rm max}$ is the frequency of the maximum of the Gaussian-like envelope the solar-like pulsations describe in the power spectrum, a better mapping of this zone implies a more accurate determination of this maximum. On the other hand, solar-like oscillations are forced oscillations of stable pulsational modes. Therefore, every individual mode has a certain lifetime. With the time, modes appear and disappear. A small $T$ will act as a picture of the living modes at this moment. They don't ensure a perfect definition of $\nu_{\rm max}$. The larger the $T$, the larger the number of observed modes, contributing to a more precise estimation of the global envelope.

Therefore, we expect a dependence of the accuracy in the determination of $\nu_{\rm max}$ with $T$.

In addition, the duty cycle produces a spurious signal in the frequency domain \citep{Moser09}, with a potential impact on the determination of $\nu_{\rm max}$. Following \citet{Stahn08}, the impact of duty cycles larger than 60 $\%$ in the determination of the position and/or power of a frequency in the Fourier domain is negligible. Nevertheless, we have studied this impact using {\it Kepler} data.

To measure the extent of these dependencies, we have simulated different realistic cases and tested the potential accuracy we can reach with {\it CHEOPS} data. To do so, we have used ten real cases as reference, displayed in Table \ref{tab:star_test}. We have analyzed a group of Short-Cadence (SC) {\it Kepler} light-curves of eight well-known Main Sequence stars and two giants. The choosing of SC is due to its similarity with {\it CHEOPS} integration time. In that way, we can compare the value of $\nu_{\rm max}$ taking into account shorter light curves and lower duty cycles with the reference ones obtained with the complete light-curves. The duty cycle of {\it CHEOPS} is strongly determined by the SAA (South Atlantic Anomaly) and also by the Earth occultation (EO). These effects depend on the orbital parameters and the position of the star during the run. We used the {\it CHEOPSim} tool (see Sect.~\ref{sec:soft}) in order to simulate several effective times of observation and the timing of the gaps at different sky positions, that is, we simulate different duty cycle values (see Table \ref{tab:duty_cycle}, column "SAA \& EO"). These simulations are superimposed on the Kepler light-curves (column "Complete"). In that way, the duty cycle of the light curves we used is lower due to the own duty cycle of the original Kepler light-curves. Finally, we interpolate these gaps with a linear fit.

We use the weighted mean frequency to calculate the frequency at maximum power \citep{Kallinger10}

\begin{equation}
\nu_{\rm max,w}=\frac{\sum A_i \nu_i}{\sum A_i}
\label{eq:kallinger}
\end{equation}

\noindent where $A_i$ and $\nu_i$ are the amplitude and frequency of each peak of the power spectra, respectively. We also define the efficiency of detection as

\begin{equation}
{\rm Eff}(\%)=100\cdot\frac{N_{\rm detections}}{N_{\rm all}}
\end{equation}

\noindent where $N_{\rm all}$ is the number of runs for a fixed length and duty cycle, taking into account all tested stars, and $N_{detections}$ is the number of significant detections inside the detection range $\nu_{\rm max,w} \in \nu_{\rm max, est}\pm 5\Delta \nu_{\rm est}$, where $\nu_{\rm max,est}$ and $\Delta \nu_{\rm est}$ are the estimated $\nu_{\rm max}$ and $\Delta \nu$ using scaling relations and the position of the star in the HR Diagram. A clear example is shown in Fig. \ref{fig:dc_1} where the relative error, defined as the relative difference between the measured and the reference $\nu_{\rm max}$ values, is calculated for several 1-day runs for one of our testing stars. We obtained values within the detection range for most of the runs (purple line). However, one single run can introduce a considerably high relative error ($\sim 34\%$). For that reason, we studied two different cases: The first case (C1) taking into account the $\nu_{\rm max, est}\pm 5\Delta \nu_{\rm est}$ range, and the second one with a shorter range (C2; $\nu_{\rm max, est}\pm 2\Delta \nu_{\rm est}$). We can use these ranges because the distribution of p-modes of the solar-oscillator power spectra around its $\nu_{\rm max}$ is approximately symmetric (they have an asymmetry of $\sim 3\%$, \citealt{Kallinger10_2}). We noted that a high accuracy of $\nu_{\rm max, est}$ and $\Delta \nu_{\rm est}$ are not required since we only need them to roughly estimate the range where $\nu_{\rm max}$ is searched.

To study the influence of the duty cycle in the determination of $\nu_{\rm max}$, we have analyzed the ten stars of Table \ref{tab:star_test}. Since for solar-like pulsations the S/N level achieved by Kepler for a particular magnitude star will be achieved by CHEOPS for a brighter star, in Table \ref{tab:star_test} we show the S/N of these testing {\it Kepler} stars and not their magnitudes. Our study is valid, therefore, when these S/N are reached. We have divided again their light-curves in runs of a fixed observational time. We have then imposed the {\it CHEOPS} duty cycle and obtained the $\nu_{\rm max}$ of every run. Finally we have obtained the mean values of the relative error in the determination of $\nu_{\rm max}$ compared with the reference value per duty cycle for all the stars and runs, the mean maximum relative error, the detection efficiency, and their standard deviations.

In the case C1, the mean relative error for observational times of one day (top left panel of Fig. \ref{fig:mean_1}) is in the range [3.8, 4.5]$\%$, depending on the duty cycle. In general, the larger the duty cycle, the lower the mean relative error. The mean maximum relative error ranges between [12.5, 15]$\%$ for an observational time of one day (top left panel of Fig. \ref{fig:max_1}). Again, the larger the duty cycle, the lower the mean maximum relative error. In terms of the detection efficiency, in the case of observational times of one day (top left panel of Fig. \ref{fig:prob_1}), this efficiency is almost stable at a value of 99.6$\%$ for every duty cycle. In the rest of the panels of Figs. \ref{fig:mean_1} to \ref{fig:prob_1} we can see the effect of increasing the observational time to 2, 4, 8 and 30 days. We noted that the length of the observations for CHEOPS will be relatively short. We include all these lengths to understand the evolution of these values with $T$ up to an observational time similar to that of {\it TESS}. In general, the larger the observational time, the lower the mean relative error, the lower the mean maximum relative error and the larger the detection efficiency. If we go into the details, the mean relative error present similar results from 4 days of observational time, on. That is, in terms of a mean relative error, when we use different measurements, the benefit of increasing the observational time is clear up to an observational time of 4 days. For larger observational times this benefit is not so justified. In terms of the mean maximum relative error, that is, the maximum error we achieve in a single run, the increasing of the observational time become in a decreasing of this mean maximum error for every observational time studied, from a range of [12.5, 15]$\%$ for one day down to a range of [2.5,4]$\%$ for 30 days of observational time. The detection efficiency is of 100$\%$ from 4 days of observational time, on.

On the other hand, in the case C2 the situation is similar to the case C1 with the following differences:

\begin{itemize}
\item The mean relative error have slightly lower values in general ([3.5, 4]$\%$ in the case of one day of observational time, for example)
\item The mean maximum error is even better, with a range of [8, 11]$\%$ in the case of one day of observational time.
\item The detection efficiency is a little worse, since C2 is more restrictive, with values always larger than 96$\%$ and with a 100$\%$ from 4 days of observational time, on.
\end{itemize}

In conclusion, although there is no large variations of the mean relative error with the duty cycle produced by the SAA or the Earth occultation, we have to take into account the possible high error of an individual measurement. Therefore, several runs are advised to discard those values of $\nu_{\rm max}$ out of range. Moreover, the highest accuracy we obtain for $\nu_{\rm max}$, the highest accuracy of $\Delta \nu$ will be achieved. Then, a proper detection range could be used, improving iteratively our results. In addition, it is not worth to propose observational ranges lager than 4 days if several runs are planned, since the impact of a maximum relative error in a single run is mitigated and no significant improvements in the mean relative error are achieved.

\section{Expected stellar uncertainties}
\label{sec:Prec}

The basic idea for determining stellar parameters is entering observational quantities such as stellar metallicity [Fe/H], effective temperature $T_{\mathrm{eff}}$ and surface gravity $\log{g}$ (usually available from spectroscopy) in a grid of theoretical models and perform a proper interpolation scheme to retrieve those quantities that best match observations. Asteroseismic $\nu_{\mathrm{max}}$ is a proxy for $\log{g}$ and its knowledge enables to better constrain the input $\log{g}$ so that to have a refinement of the output stellar parameters.
In fact, once the asteroseismic $\log{g}$ is recovered through Eq. \ref{eq:numax}, spectroscopic and asteroseismic $\log{g}$ can be combined in a weighted mean to obtain a better estimation of the stellar surface gravity and to decrease its uncertainty. 

If $\nu_{\mathrm{max}}$ is added among the input parameters, in this section we want to:
\begin{enumerate}
 \item give a reasonable estimate of the precision we gain in the input $\log{g}$;
 \item test the precision we would gain in the output parameters.
\end{enumerate}

Given several measurements $g_i$ of an observable $g$, whose uncertainties are $\sigma_i$, the weighted mean is computed as
\begin{equation}
 \bar{g}=\frac{\sum_i g_i w_i}{\sum_i w_i}
 \label{eq:wmean}
\end{equation}
where the weight $w_i=\sigma_i^{-2}$, and its uncertainty is
\begin{equation}
 \sigma_{\bar{g}}=\frac{1}{\sqrt{\sum_i w_i}}
 \label{eq:Uwmean}
\end{equation}

Error propagation from Eq. \ref{eq:numax} suggests that the relative uncertainty on the derived surface gravity is
\begin{equation}
 \frac{\Delta g}{g}=\frac{\Delta \nu_{\mathrm{max}}}{\nu_{\mathrm{max}}}+\frac{1}{2}\frac{\Delta T_{\mathrm{eff}}}{T_{\mathrm{eff}}}
 \label{eq:Deltagg}
\end{equation}

Re-writing Eqs. \ref{eq:wmean} and \ref{eq:Uwmean} in terms of relative uncertainties $\delta_i=\sigma_i/g_i$, the relative uncertainty on the weighted mean $\delta_{\bar{g}}$ results to be
\begin{equation}
 \delta_{\bar{g}}=\frac{\sigma_{\bar{g}}}{\bar{g}}=\frac{\sqrt{\displaystyle\sum_{i=1}^n \frac{1}{g_i^2 \delta_i^2}}}{\displaystyle\sum_{i=1}^n \frac{1}{g_i \delta_i^2}}
 \label{eq:dwmean}
\end{equation}
where $n$ is the number of available measurements. If $n=2$ (as in our case), Eq. \ref{eq:dwmean} becomes
\begin{equation}
 \delta_{\bar{g}}=\delta_1 \delta_2 \frac{\sqrt{g_1^2\delta_1^2+g_2^2\delta_2^2}}{g_1\delta_1^2+g_2\delta_2^2} = \delta_1 \delta_2 \frac{\sqrt{\delta_1^2+k^2\delta_2^2}}{\delta_1^2+k\delta_2^2}
 \label{eq:dwmean2}
\end{equation}
where $k=g_2/g_1$. Studying $\delta_{\bar{g}}$ as a function of $k$, it turns out that it has an absolute minimum for $k=1$ i.e. when $g_1=g_2$ and that $\lim_{k\to 0} \delta_{\bar{g}}(k)=\delta_2$ (i.e. when $g_1\gg g_2$), $\lim_{k\to +\infty} \delta_{\bar{g}}(k)=\delta_1$ (i.e. when $g_2\gg g_1$).

Identifying spectroscopic data with the subscript 1 and asteroseismic data with the subscript 2, it turns out that the median relative uncertainty of the spectroscopic surface gravity of the ensemble of {\it CHEOPS} reference targets (it's made of 152 stars and their main properties are listed in Table \ref{tab:CHEOPS_sampl}) is $\bar{\delta_1}\approx0.21$, while, instead,  $\bar{\delta_2}\approx0.055\approx\frac{\Delta\nu_{\mathrm{max}}}{\nu_{\mathrm{max}}}$, assuming $5\%$ as a conservative estimation of the uncertainty on ${\nu_{\mathrm{max}}}$ as it comes out from Sect.~\ref{sec:Acc} and considering the weak dependency of Eq. \ref{eq:Deltagg} on the relative uncertainty on $T_{\mathrm{eff}}$. Thus, as a median value of reference, $\bar{\delta_1}\approx4\bar{\delta_2}$. If this relation holds and if we consider a possible difference of up to a factor of 2 between the two surface gravity determinations $g_1$ and $g_2$, we obtain $\delta_{\bar{g}}(k=\frac{1}{2})\approx0.98\delta_2$ and $\delta_{\bar{g}}(k=2)\approx0.99\delta_2$ (the minimum is $\delta_{\bar{g}}(k=1)\approx0.97\delta_2$).
This statistical overview suggests that, for a star by star analysis, the maximum value between $\delta_{\bar{g}}(k=\frac{1}{2})$ and $\delta_{\bar{g}}(k=2)$
\begin{equation}
 \delta_g^*=\mathrm{max}(\delta_{\bar{g}}(k=1/2), \delta_{\bar{g}}(k=2))
 \label{eq:deltag}
\end{equation}
is a reasonably conservative estimate of the relative uncertainty of the weighted mean and this estimate is likely lower than $\delta_2$.

\begin{table*}
	\centering
	\caption{Summary of the sampling of {\it CHEOPS} stars.}
	\begin{tabular}{ccc}
		\hline
		Charact. & Min & Max \\
		\hline
		$T_{\rm eff}$ ($^\circ$K)&3030 & 10200 \\
		log$L$ (dex)& -2.38& 1.72\\
		log$g$ (dex)& 3.47 & 5.05\\
		Mass (M$_\odot$) & 0.2 & 2.71\\
		\hline
	\end{tabular}
	\label{tab:CHEOPS_sampl}
\end{table*}

To achieve the second goal, we considered the {\it CHEOPS} sample specified in Table \ref{tab:CHEOPS_sampl}. The estimate of stellar output parameters has been done thanks to the Isochrone Placement algorithm described in \citet{bonfanti15,bonfanti16}. Interpolation in theoretical grids of tracks and isochrones have been made considering PARSEC\footnote{%
Padova and Trieste Stellar Evolutionary Code. http://stev.oapd.inaf.it/cgi-bin/cmd}%
 evolutionary models, version 1.2S (see \citealt{bressan12,chen14}; and references therein).

\begin{table}
	\caption{Relative uncertainties expressed in \% on the age $t$, the mass $M$, the radius $R$ and the mean stellar density $\rho$ for the different runs. See text for details.}
	\label{tab:relUncertainties}
	\begin{tabular}{ccc}
		\toprule
		&	Input param.	& {\it CHEOPS} sample		\\
		\midrule
		\multirow{4}*{$\frac{\Delta t}{t}$}&standard					& 52	\\
		&with $\nu_{\mathrm{max,best}}$			& 38	\\
		&with $0.9\nu_{\mathrm{max,best}}$		& 47	\\
		&with $1.1\nu_{\mathrm{max,best}}$		& 48	\\
		\midrule
		\multirow{4}*{$\frac{\Delta M}{M}$}&standard					& 2.1	\\
		&with $\nu_{\mathrm{max,best}}$  			& 1.8	\\
		&with $0.9\nu_{\mathrm{max,best}}$		& 2.2	\\
		&with $1.1\nu_{\mathrm{max,best}}$		& 2.1	\\
		\midrule
		\multirow{4}*{$\frac{\Delta R}{R}$}&standard					& 1.8	\\
		&with $\nu_{\mathrm{max,best}}$  			& 1.6	\\
		&with $0.9\nu_{\mathrm{max,best}}$		& 1.9	\\
		&with $1.1\nu_{\mathrm{max,best}}$		& 1.9	\\
		\midrule
		\multirow{4}*{$\frac{\Delta\rho}{\rho}$}&standard				& 5.6	\\
		&with $\nu_{\mathrm{max,best}}$  			& 4.7	\\
		&with $0.9\nu_{\mathrm{max,best}}$		& 5.7	\\
		&with $1.1\nu_{\mathrm{max,best}}$		& 5.6	\\
		\bottomrule
	\end{tabular}
\end{table}

The code has been run four times. The first time, the input parameters were [Fe/H], $T_{\mathrm{eff}}$, $\log{g}$ coming from spectroscopy, $v\sin{i}$ and/or $\log{R'_{\mathrm{HK}}}$ where available to improve convergence during interpolation, and parallax $\pi$ and mean $G$ magnitude coming from Gaia DR2 archive\footnote{%
https://gea.esac.esa.int/archive/} %
so that to have a measure of the stellar luminosity $L$. If data from Gaia were not available (in 4 cases), parallaxes have been taken from Hipparcos \citep{vanLeeuwen07}. We will refer to this set of input data as standard input parameters. At the end of all the cross-matches to retrieve the input parameters, our reference testing sample coming from Cheops is made of 143 stars. 
The second time, we also wanted to take the contribution from asteroseismology into account. No $\nu_{\mathrm{max}}$ values are available for the {\it CHEOPS} targets so far, therefore, for each star of the sample, we generated a set of possible $\nu_{\mathrm{max}}$ values. According to the scaling relation (Eq. \ref{eq:numax}), we computed $\nu_{\mathrm{max,input}}$ and its uncertainty considering the input values of $T_{\mathrm{eff}}$ and $\log{g}$, so that to establish the maximum range [$\nu_{\mathrm{max,input}}-\Delta\nu_{\mathrm{max,input}}$, $\nu_{\mathrm{max,input}}+\Delta\nu_{\mathrm{max,input}}$] of plausible variation of $\nu_{\mathrm{max}}$, consistently with the other input parameters. Starting from the left-most side value of the interval, we generated a sequence of $\nu_{\mathrm{max,}i}$ such that they belong to the interval and whose relative uncertainty was $5\%$.
Each value of the sequence has been obtained by adding to the previous one half of its error bar. This second run of the algorithm involves lots of 'fake' stars because of the arbitrary choice of the $\nu_{\mathrm{max,}i}$. Therefore, among all the results, for each star we considered that set of output parameters that match the theoretical models best, which derive from a specific $\nu_{\mathrm{max,best}}$ value. So, the results we will consider in this case derive from the standard input parameters and the $\nu_{\mathrm{max,best}}$. These results represent the best theoretical improvement that is expected for stars located in that region of the HRD once that $\nu_{\mathrm{max}}$ is added in input.
Finally, the third and fourth case consider the standard input parameters plus a value of $\nu_{\mathrm{max}}$ that has been obtained by decreasing and increasing $\nu_{\mathrm{max,best}}$ by $10\%$. In fact, we want also to test whether adding an additional input parameter always determine an improvement in the output uncertainties, regardless of its nominal value.

After that, we compared the relative uncertainties affecting the output age, mass, radius and mean stellar density in the four runs. The results are synthesized in Tab. \ref{tab:relUncertainties}. In this table we show the median relative uncertainties, a robust indicator against outliers. In general, the inclusion of $\nu_{\mathrm{max}}$ as an additional observable improves the precision in the determination of the stellar mass, radius, density and age. The amount of this improvement depends on the sampling used and the output analyzed.
On the one hand, when the used $\nu_{\rm max}$ is fully consistent with the selection obtained using the standard inputs (as it is the case $\nu_{\rm max,best}$), then age uncertainties are reduced from 52$\%$ to 38$\%$; mass uncertainties from 2.1$\%$ to 1.8$\%$; radius uncertainties from 1.8$\%$ to 1.6$\%$; density uncertainties from $5.6\%$ to $4.7\%$. Repeating this entire analysis without the input Gaia luminosity (which already gives a strong and straightforward constraint on the radius), it turned out that the benefit of adding $\nu_{\mathrm{max}}$ is even more sensible in reducing the output uncertainties of $R$ and $M$.
On the other hand, when the consistency between the standard selection and $\nu_{\rm max}$ is deteriorated, the improvement is also deteriorated. This simulation shows that both the input precision and the consistency among the input parameters play a role in reducing the median output uncertainties.

It is worth to add that also the star location on the HRD influences the improvement level in the output uncertainties if we add further input parameter. For instance, if the evolutionary stage of a star is around and soon after the turn-off (TO), there theoretical models are very well spaced, a star can be easily characterized and adding further input parameters don't make change things that much. Instead, if a star is still on the main sequence (MS) or it is well evolved after the TO, there theoretical models are very close and the reduction in the output uncertainties when an input parameter is added may be remarkable. To prove these considerations, we have analyzed the Kepler stars of Table \ref{tab:star_test}, that are almost all around the TO region. We have used the standard constraints $T_{\rm eff}$, log$g$, [Fe/H], $L$ (and $v\sin{i}$ where available) on the one hand, and we have added the observed $\nu_{\rm max}$ with an uncertainty of a 5$\%$ as a conservative maximum uncertainty derived from Sect.~\ref{sec:Acc}. We have artificially homogenized all the log$g$ uncertainties to a minimum value of 0.1 dex. to reproduce what we usually obtain from spectroscopy. In a median sense, no relevant variations in the output uncertainties is seen adding $\nu_{\mathrm{max}}$ among the input. Besides the fact that here the consistency between the subgroup fitting the standard observations and the observed $\nu_{\rm max}$ is not ensured, the majority of these stars are located around the TO where isochrones are well spaced and Gaia luminosity already provides a precise location for the stars.
But if we move on a star-by-star analysis, this Kepler sample contains two stars that are well evolved (i.e. located strongly beyond the TO), namely KIC 5701829 and KIC 9145955. Adding $\nu_{\mathrm{max}}$ as a further input provokes a reduction in the output parameters of these stars as reported in the following:
\begin{itemize}
\item KIC 5701829. $\frac{\Delta t}{t}$ from 45\% to 18\%; $\frac{\Delta M}{M}$ from 12\% to 7\%; $\frac{\Delta R}{R}$ from 3.3\% to 2.8\%; $\frac{\Delta\rho}{\rho}$ from 18\% to 11\%.
\item KIC 9145955. $\frac{\Delta t}{t}$ from 46\% to 38\%; $\frac{\Delta M}{M}$ from 13\% to 10\%; $\frac{\Delta R}{R}$ from 4.1\% to 3.2\%; $\frac{\Delta\rho}{\rho}$ from 23\% to 18\%.
\end{itemize}
We can conclude that the reductions can be sensible; the improvement in the $R$ precision is less evident just because we already have a precise knowledge of $R$ thanks to Gaia.


As an important final remark, we stress that all the relative uncertainties we have provided are internal at 1-$\sigma$ level originating from the interpolation scheme in theoretical models. The statistical treatment, the density of the model grids in use, how the uncertainty on [Fe/H] has been addressed, the treatment of the element diffusion enter a complicated picture and have all a role in affecting the output uncertainties. What is relevant here is judging the level of improvement on the output depending on the input set. Moreover let us note that all these results are for a given input physics (e.g. opacities, equation of state, nuclear reaction rates) and a given initial He abundance (that cannot be determined from spectroscopy) in stellar theoretical models. Constraining input physics and/or initial He abundance from asteroseismology requires very high quality individual oscillation frequencies of the star considered, such as those provided by \textit{CoRoT} or \textit{Kepler} space missions (e.g. \citealt{2014A&A...569A..21L,2016A&A...585A.109B,2016A&A...596A..73B}).

\section{Conclusions}
\label{sec:concl}

In this work, we have studied the asteroseismic potential of {\it CHEOPS}. We have found that, with the current instrumental performance and observational times between eight hours and two days, the asteroseismic observable $\nu_{\rm max}$ can be determined for massive F and G-type stars from late MS on, and for all F, G, and K-type stars from post-MS on. This observational times perfectly fit the observational strategy of {\it CHEOPS}.

The estimated $\nu_{\rm max}$ accuracy, obtained using ten {\it Kepler} light-curves as reference, is of the order of 5$\%$ or better when the star is observed several times, independently of the expected duty cycles of the {\it CHEOPS} targets. In addition, the larger the observational time, the larger the HR Diagram zone where the $\nu_{\rm max}$ can be detected and the better the accuracy.

This accuracy in the determination of $\nu_{\rm max}$ is translated into a similar precision in the determination of $\log{g}$, which is around four times smaller than the precision obtained from spectroscopy, in median values. This new precision and the inclusion of an additional observable for fitting the theoretical models reduce the uncertainty in the determination of the stellar mass, radius, and age depending on the stellar location on the HRD and on the degree of consistency between the expected $\nu_{\rm max}$ obtained using standard inputs and the observed one. Having a complete set of spectroscopic input parameters plus the stellar luminosity from Gaia, our theoretical simulation on a testing sample of {\it CHEOPS} stars shows that, once a consistent $\nu_{\mathrm{max}}$ is available in input, in median sense age uncertainty decreases from 52\% to 38\%, mass uncertainty from 2.1\% to 1.8\%, radius uncertainty from 1.8\% to 1.6\% and density uncertainty from 5.6\% to 4.7\%. All these provided uncertainties are meant to be internal.

In addition, we have also found that the {\it CHEOPS} light curves can provide an accurate estimation of the $\delta$~Scuti $\nu_{\rm max}$ (see Appendix), leading to a measurement of the stellar rotational rate, its inclination with respect the line of sight, and its mean effective temperature.

This work open an opportunity for complementing {\it TESS} asteroseismic observations since {\it CHEOPS} has its technical strengths where {\it TESS} has some weaknesses: stars in the ecliptic plane and/or fainter than {\it TESS} limit thanks to its larger telescope aperture.


%

\begin{acknowledgements}
      The authors want to acknowledge the anonymous referee for his/her valuable and constructive comments, and William J. Chaplin for very fruitful discussions. AM acknowledges funding from the European Union's Horizon 2020 research and innovation program under the Marie Sklodowska-Curie grant agreement No 749962 (project THOT). A.M., S.B.F and D.B. acknowledge support by the Spanish State Research Agency (AEI) Project No. ESP2017-87676-C5-1-R. S.B.F. and D.B. acknowledge support by the Spanish State Research Agency (AEI) Project No. MDM-2017-0737 Unidad de Excelencia “María de Maeztu”- Centro de Astrobiología (INTA-CSIC). V.V.G. is F.R.S.-FNRS Research Associate. A.B. is funded by an Action de Recherche Concert\'ee (ARC) grant financed by the Wallonia-Brussels Federation. S.J.A.J. is funded by the CHEOPS Prodex Grant PEA 4000113509. Caph results are based on observations made with the Hertzsprung SONG telescope operated on the Spanish Observatorio del Teide on the island of Tenerife by the Aarhus and Copenhagen Universities and by the Instituto de Astrof\'isica de Canarias. This research has made use of the Exoplanet Orbit Database and the Exoplanet Data Explorer at exoplanets.org.

\end{acknowledgements}

\section*{Software}
\label{sec:soft}

The CHEOPS instrument and science simulator, \texttt{CHEOPSim}, is developed under the responsibility of the CHEOPS Mission Consortium. \texttt{CHEOPSim} is implemented by D.~Futyan as part of the Science Operation Centre located at the University of Geneva. Analysis was performed with R version 3.3.1 \citep{R}, RStudio Version 1.0.143, and the R libraries dplyr 0.5.0 \citep{dplyr2016}, parallel \citep{R}, RcppParallel \citep{RcppParallel}, rdetools \citep{rdetools}, plor3D \citep{plot3D}. The non-linear fitting method used to interpolate $\delta$~Scuti light curves is part of the $\delta$~Scuti Basics Finder pipeline \citep[$\delta$SBF;][and references therein]{Barcelo17} 


\begin{appendix}
\section{Influence of the duty cycle in $\nu_{\rm max}$ determination for $\delta$~Scuti stars}
\label{sec:scuti}

\begin{figure*}
\resizebox{\hsize}{!}{\includegraphics[width=\columnwidth]{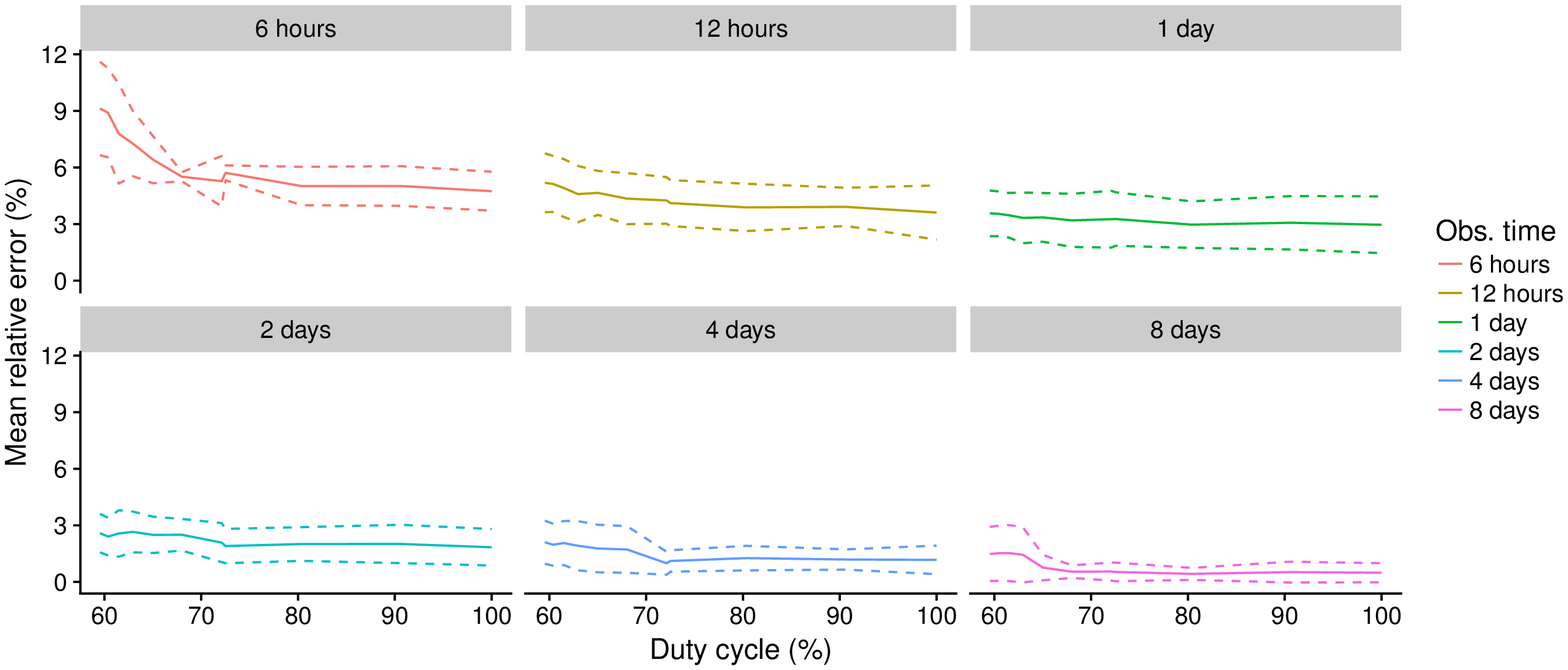}}
 \caption{Mean relative error of $\nu_{\rm max}$ with duty cycle for the tested $\delta$~Scuti stars. The different colours point to a different length of the run 6 and 12 hours, 1, 2, 4 and 8 days. Dashed lines are $\pm$ their standard deviation}
 \label{fig:Mean_delta}
\end{figure*}

\begin{figure*}
\resizebox{\hsize}{!}{\includegraphics[width=\columnwidth]{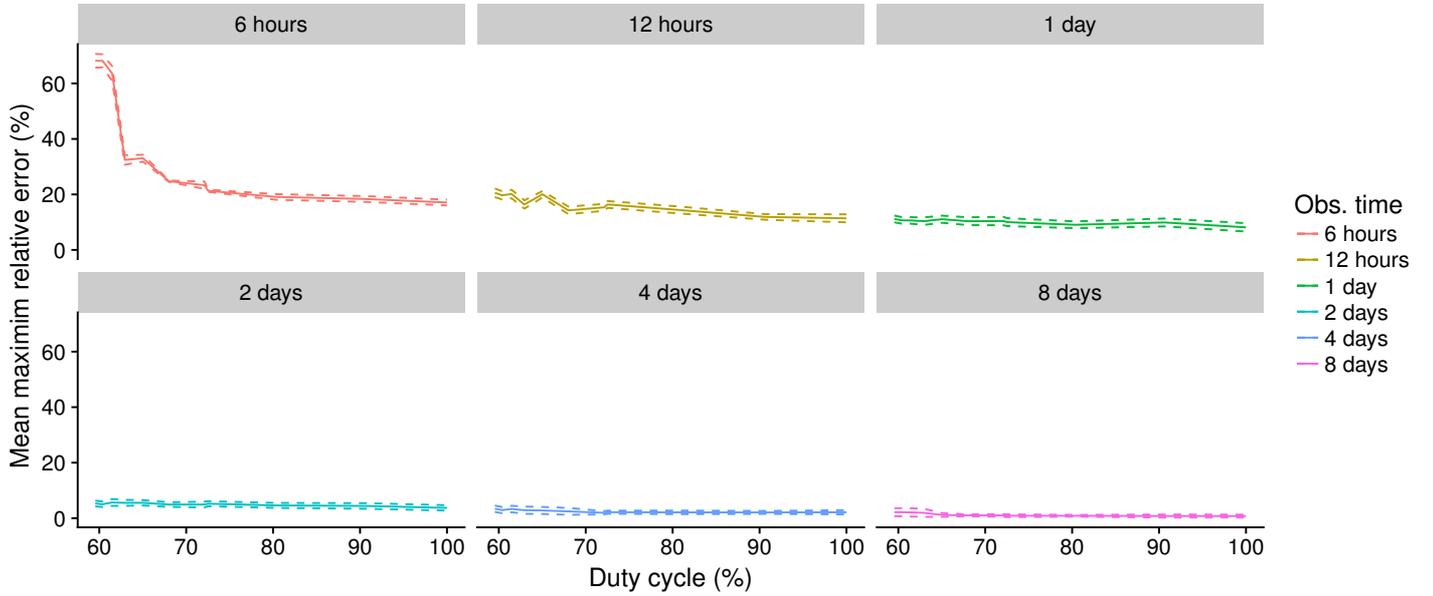}}
 \caption{Mean maximum relative error of $\nu_{\rm max}$ with duty cycle for the tested $\delta$~Scuti stars. The different colours point to a different length of the run 6 and 12 hours, 1, 2, 4 and 8 days. Dashed lines are $\pm$ their standard deviation}
 \label{fig:Max_delta}
\end{figure*}

{\it CHEOPS} will not observe only solar-like pulsators looking for exoplanets. $\delta$~Scuti stars are interesting stellar bodies that can also present transiting exoplanets \citep[e.g.,][]{Christian06}. $\delta$~Scuti stars are classical pulsators excited by $\kappa$-mechanism \citep{Chevalier71}. Althought their pulsation frequency range is not in the solar-like regime, we can define always $\nu_{\rm max}$ using Equation~\ref{eq:kallinger} \citep[see][]{Barcelo18}.

A significant rotation can produce the oblateness of the star and a gradient of temperature from the poles to the equator known as the gravity-darkening effect \citep{Zeipel24}. \citet{Barcelo18} suggest a direct relation between the $\nu_{\rm max}$ for $\delta$~Scuti stars and their mean effective temperature ($\bar{T}_{\rm eff}$) due to its oblateness. Then, once measured the temperature with Str\"omgren photometry ($T_{\rm eff}$), we can compare it with the mean effective temperature. The relative difference will constrain the rotation rate ($\Omega/\Omega_{K}$) and the inclination with respect to the line of sight \citep{Barcelo18}.

Since $\kappa$-mechanism produces waves with higher lifetime than stochastic mechanism, their pulsations are easier to observe in short light curves. In fact, 1.6-days WIRE observations of Caph ($\beta$~Cas) prove that it is possible to detect the main oscillation frequencies from such a short light curves \citep[$\nu_{0}=115$ $\mu$Hz,][]{Cuypers02}. Our own 10 hours of ground-based observations using SONG-OT allowed us to detect its highest amplitude oscillation with only a 2\% of relative error in frequency ($\nu_{0}=117$ $\mu$Hz).

\begin{table}
	\centering
	\caption{Frequency at maximum power of the 6 $\delta$~Scuti stars.}
	\begin{tabular}{c c}
		\hline
		KIC &  $\nu_{\rm max}^a$ \\
		&  in $\mu$Hz \\
		\hline
		4374812  & 126 $\pm$ 7\\
		4847371  & 300 $\pm$ 19\\
		4847411  & 296 $\pm$ 14\\
		6844024  & 177 $\pm$ 8\\
		9072011  & 90  $\pm$ 3\\
		11285767 & 241 $\pm$ 15\\
		\hline
	\end{tabular}
	\label{tab:6dscu}
	\tablefoot{These values have been calculated as indicated in \citet{Barcelo18}.}
\end{table}

We repeated the study of the duty cycle effect on $\nu_{\rm max}$ determination for $\delta$~Scuti stars analyzing 6 Short-Cadence (SC) \textit{Kepler} light curves of A/F stars with $\delta$~Scuti pulsations (see Table~\ref{tab:6dscu}). We obtained their "true" frequencies at maximum power taking into account their entire light curve. In addition, we interpolated the gaps of each shortened light curve using a non-linear fitting method \citep{Barcelo15}. The mean relative error obtained is around 5\% for only 6 hours of run time and duty cycles higher than 70\% (see Fig.~\ref{fig:Mean_delta}). For lower duty cycles the mean relative error increases up to 9\%. However, in some shortened light curves, we can obtain a higher departure from the $\nu_{\rm max}$ value (from $\sim$20\% to $\sim$60\% depending on duty cycle). Then, like in the solar-like oscillators, it is important to make several revisits to the same star. For all the other lengths, the mean relative error does not significantly change with the duty cycle thanks to the interpolation method. Moreover, the higher the length, the lower maximum relative error (Fig.~\ref{fig:Max_delta}). Therefore, we can obtain a high accuracy in $\nu_{\rm max}$ for a $\delta$~Scuti star with light curves that contain only a few periods of the oscillation. In addition, it is guaranteed a relative error lower than 10\% for 1-day light curves and 5\% for 2-day or longer light curves.

In conclusion, observing $\delta$~Scuti stars with {\it CHEOPS} will allow us to obtain one of its seismic indices, $\nu_{\rm max}$, with high accuracy and only investing a few hours. This detection can provide a determination of the stellar rotation rate, its inclination with respect the line of sight, and its mean effective temperature.

\end{appendix}
\end{document}